\documentclass[twocolumn,preprintnumbers,amsmath,amssymb,superscriptaddress]{revtex4-1}

\usepackage{graphicx}
\usepackage{dcolumn}
\usepackage{bm}

\makeatletter
\def\@seccntformat#1{}
\makeatother
\renewcommand{\numberline}[1]{}

\begin{document}

\title{Experimental Discovery of Topological Surface States - A New Type of 2D Electron Systems}

\author{M. Zahid Hasan}
\affiliation{Joseph Henry Laboratories, Department of Physics, Princeton
University, Princeton, NJ 08544, USA}
\affiliation{Princeton Institute for the Science and Technology of Materials, School of Engineering and Applied Science, Princeton University, Princeton NJ 08544,
USA}
\author{Su-Yang Xu}
\affiliation{Joseph Henry Laboratories, Department of Physics, Princeton University, Princeton, NJ 08544, USA}
\affiliation{Advanced Light Source, Lawrence Berkeley National Laboratory, Berkeley, California 94305, USA}

\author{David Hsieh}
\affiliation{Joseph Henry Laboratories, Department of Physics, Princeton
University, Princeton, NJ 08544, USA}
\affiliation{Department of Physics, California Institute of Technology, Pasadena, CA 91125, USA}
\author{L. Andrew Wray}
\affiliation{Joseph Henry Laboratories, Department of Physics, Princeton
University, Princeton, NJ 08544, USA}
\affiliation{Advanced Light Source, Lawrence Berkeley National Laboratory, Berkeley, California 94305, USA}
\author{Yuqi Xia}
\affiliation{Joseph Henry Laboratories, Department of Physics, Princeton
University, Princeton, NJ 08544, USA}

\begin{abstract}

Topological Surface States (TSS) represent new types of two dimensional electron systems with novel and unprecedented properties distinct from any quantum Hall-like or spin-Hall effects. Their Z$_2$  topological order can be realized at room temperatures without magnetic fields and they can be turned into magnets, exotic superconductors or Kondo insulators leading to worldwide interest and activity in the topic. We review the basic concepts defining such topological matter and the experimental discovery via the key experimental probe (spin-ARPES) that revealed the Z$_2$ topological order in the bulk of these spin-orbit interaction dominated insulators for the first time. This review focuses on the key results that demonstrated the fundamental Z$_2$ topological properties such as spin-momentum locking, non-trivial Berrys phases, mirror Chern number, absence of backscattering, bulk-boundary correspondence (topology), protection by time-reversal and other discrete (mirror) symmetries and their remarkable persistence up to the room temperature elaborating on results first briefly discussed in an early review by M.Z. Hasan and C.L. Kane [Rev. of Mod. Phys., 82, 3045 (2010) ]. Additionally, key results on broken symmetry phases such as quantum magnetism and superconductivity induced in topological materials are briefly discussed. Topological insulators beyond the Z$_2$ classification such as Topological Crystalline Insulators (TCI) are discussed based on their spin properties (mirror Chern invariants). The experimental methodologies detailed here have been used in most of the subsequent studies of Z$_2$ topological physics in almost all bulk topological insulator materials to this date.

\end{abstract}


\maketitle

\tableofcontents

\section{Introduction}

The three-dimensional topological insulator (originally called ``topological insulators" to distinguish this novel state from all types of quantum Hall and spin Hall-like effects since the original quantum Hall state is the time-reversal breaking 2D topological insulator first discovered in 1980) is the first (and only known experimental) example in nature of a topologically ordered electronic phase existing in bulk solids. Their topological order can be realized at room temperatures without magnetic fields and they can be turned into magnets and exotic superconductors leading to world-wide interest and activity in topological insulators \cite{RMP, Moore1, HasanMoore, zhang, Roy, FuKM, 15, Hsieh1}. All of the 2D topological insulator examples (Integer Quantum Hall(IQH), Quantum Spin Hall(QSH)) including the fractional one (FQH) involving Coulomb interaction are understood in the standard picture of quantized electron orbits in a spin-independent or spin-dependent magnetic field, the 3D topological insulator defies such description and is a novel type of topological order which cannot be reduced to multiple copies of quantum-Hall-like states. In fact, the 3D topological insulator exists not only in zero magnetic field but differs from the 2D variety in three very important aspects: \textbf{1}) they possess topologically protected 2D metallic surfaces (Topological Surface States, a new type of 2DEG) rather than the 1D edges, \textbf{2}) they can work at room temperature (300K and beyond) rather than cryogenic (sub-K) temperatures required for the QSH effects and \textbf{3}) they occur in standard bulk semiconductors rather than at buried interfaces of ultraclean semiconductor heterostructures and thus tolerate stronger disorder than the IQH-like states. One of the major challenges in going from quantum Hall-like 2D states to 3D topological insulators is to employ new experimental approaches/methods to precisely probe this novel form of topological-order since the standard tools and settings that work for IQH-states also work for QSH states. The method to probe 2D topological-order is exclusively with charge transport. In a 3D topological insulator, the boundary itself supports a 2DEG and transport is not (Z$_2$) topologically quantized hence \textit{cannot} directly probe the topological invariants {$\nu_o$} nor the topological quantum numbers analogous to the Chern numbers of the IQH systems. This is \textit{unrelated} to the fact that the present materials have some extrinsic or residual/impurity conductivity in their naturally grown bulk. In this paper, we review the birth of momentum- and spin-resolved spectroscopy as a new experimental approach and as a highly boundary sensitive method to study and prove topological-order via the direct measurements of the topological invariants {$\nu_o$} that are associated with the Z$_2$ topology of the spin-orbit band structure and opposite parity band inversions. These experimental methods led to the experimental discovery of the first 3D topological insulator in Bi-Sb semiconductors which further led to the discovery of the Bi$_{2}$Se$_3$ class - the most widely researched topological insulator to this date. We discuss the fundamental properties of the novel topologically spin-momentum locked half Dirac metal on the surfaces of the topological insulators and how they emerge from topological phase transitions due to increasing spin-orbit coupling in the bulk. These methods and their derivatives are now being applied by others world-wide for further finer investigations of topological-order and for discovering new topological insulator states as well as exotic topological quantum phenomena. We also review how spectroscopic methods are leading to the identification of spin-orbit superconductors that may work as Majorana platforms and can be used to identify topological superconductors - yet another class of new states of matter.

\section{The birth of momentum-resolved spectroscopy as a direct experimental probe of Z$_2$ Topological-Order}

Ordered phases of matter such as a superfluid or a ferromagnet are
usually associated with the breaking of a symmetry and are
characterized by a local order parameter \cite{1}. The typical
experimental probes of these systems are sensitive to order
parameters. In the 1980s, two new phases of matter were realized by subjecting 2D electron gases at buried interfaces of semiconductor heterostructures to large magnetic fields. These new phases of matter, the 2D integer and 2D fractional quantum Hall states, exhibited a new and rare type of order that is derived from an
organized collective quantum entangled motion of electrons \cite{2,3,4,Tsui}. These
so-called ``2D topologically ordered insulators" do not exhibit any
symmetry breaking and are characterized by a topological number \cite{5}
as opposed to a local order parameter. The most striking manifestation of this 2D topological order is the existence of one-way propagating 1D metallic states confined to their edges, which lead to remarkable quantized charge transport phenomena. To date the experimental
probe of their topological quantum numbers is based on charge transport, where measurements of the quantization of transverse magneto-conductivity $\sigma_{xy} =
ne^{2}/h$ (where $e$ is the electric charge and $h$ is Planck's
constant) reveals the value of the topological number $n$ that
characterizes these quantum Hall states \cite{6}.

\begin{figure*}
\includegraphics[scale=0.55,clip=true, viewport=0.0in 0.0in 11.5in 8.5in]{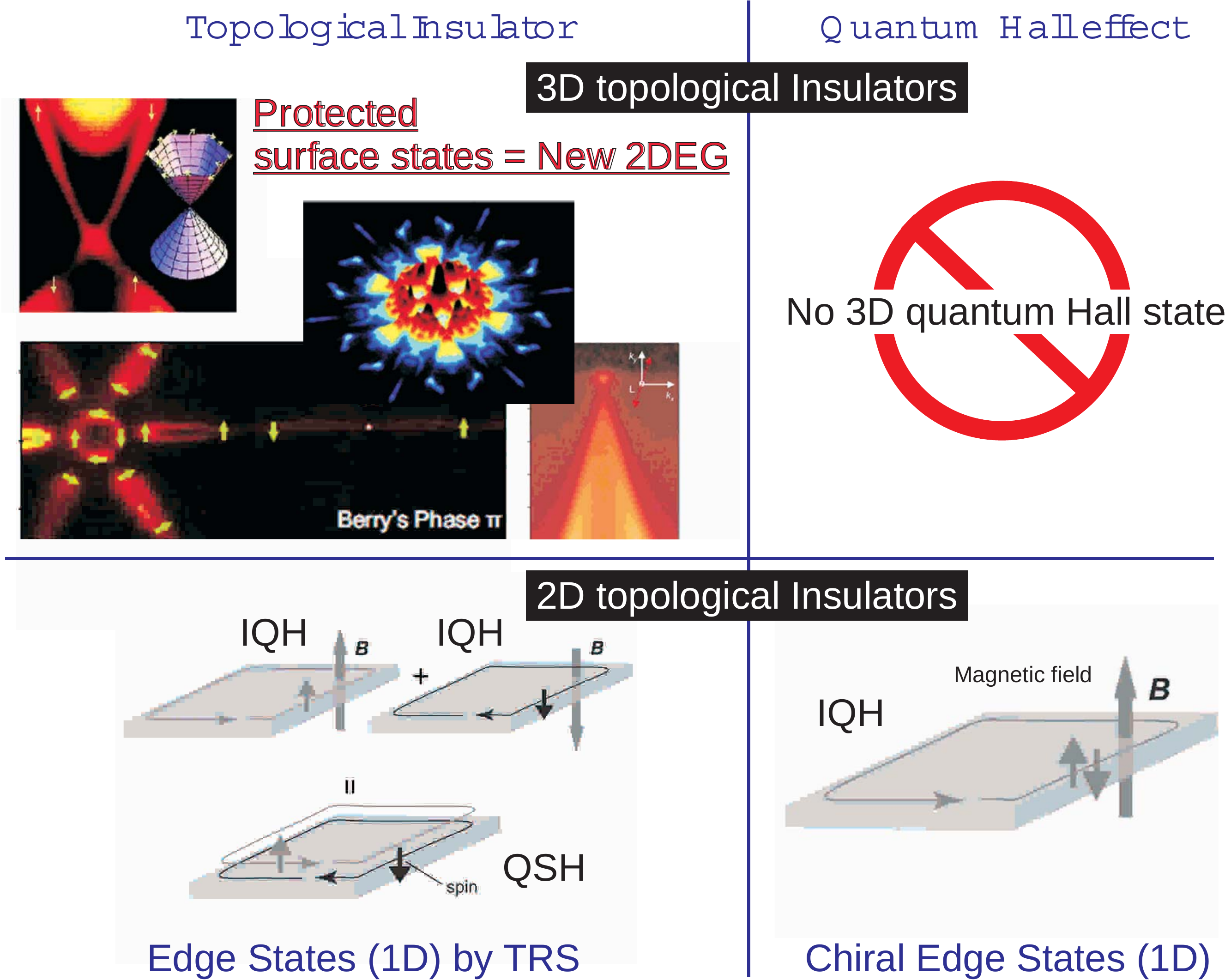}
\caption{\label{Intro} \textbf{2D and 3D topological insulators in nature.} The 2D topological insulator class consists of the 2D quantum Hall states (IQH) and 2D quantum spin Hall state (QSH). The latter is constructed from two copies of the former. On the other hand, a 3D quantum Hall state is forbidden in nature, so the 3D topological insulator represents a new type of topologically ordered phase. The protected surface states form a novel type of topological metal (half Dirac metal) where electron's spin is locked to its momentum but exhibit no spin quantum Hall effect \cite{RMP, Moore1, HasanMoore, zhang, Roy, FuKM, 15, Hsieh1}}
\end{figure*}

Recently, a third type of 2D topological insulator, the spin quantum Hall insulator, was theoretically predicted \cite{14,8} and then experimentally discovered \cite{7}. This class of quantum Hall-like topological phases can exist in spin-orbit materials without external magnetic fields, and can be described as an ordinary quantum Hall state in a spin-dependent magnetic field. Their topological order gives rise to counter-propagating 1D edge states that carry opposite spin polarization, often described as a superposition of a spin up and spin down quantum Hall edge state (Figure 1). Like conventional quantum Hall systems, the 2D spin quantum Hall insulator (QSH) is realized at a buried solid interface. The first realization of this phase was made in (Hg,Cd)Te quantum wells using \textit{charge transport} by measuring a longitudinal conductance of about $2e^2/h$ at mK temperatures \cite{7}. The quantum spin Hall state can be thought of as two copies of integer quantum Hall states (IQH) and are protected by a Z$_2$ invariant.

It was also realized that a fundamentally new type of genuinely three-dimensional topological order might be realized in bulk crystals without need for an external magnetic field \cite{Fu:STI2,11,15}. Such a 3D topological insulator cannot be reduced to multiple copies of the IQH and such phases would be only the fourth type of topologically ordered phase to be discovered in nature, and the first type to fall outside the quantum Hall-like 2D topological states (IQH, FQH, QSH). Instead of having quantum-Hall type 1D edge states, these so-called 3D topological insulators would possess unconventional metallic 2D topological surface states called spin-textured helical metals, a type of 2D electron gas long thought to be impossible to realize. However, it was recognized that 3D topological insulators would NOT necessarily exhibit a topologically (Z$_2$) quantized charge transport by themselves as carried out in the conventional transport settings of all quantum-Hall-like measurements. Therefore, their 3D topological quantum numbers (Z$_2$), the analogues of $n$ (Chern numbers), could not be measured via the charge transport based methods even if a complete isolation of surface charge transport becomes routinely possible. Owing to the 2D nature of the two surface conduction channels (top and bottom surfaces of a typical thin sample) that contribute together in a 3D topological insulator case, it was theoretically recognized that it would not be possible to measure the topological invariants due to the lack of a quantized transport response of the 2D surface that measures the Z$_2$ topological invariants  \cite{Fu:STI2}.

Here we review the development of spin- and angle-resolved photoemission spectroscopy (spin-ARPES) as the new approach/method to probe 3D topological order \cite{10,Science, Xia, Zhang_nphys, Nature_2009, Chen, Hsieh_PRL, Wray1, Ternary arXiv, Heusler, Hsin Thallium, Wray2, Xu, Ternary PRB, Hedgehog, TCI Hasan}, which today constitutes the experimental standard for identifying topological order in bulk solids. 3D topological insulators are also experimentally studied by many others world-wide using various techniques such as ARPES \cite{Xue Nature physics QL, Chen Science Fe, Sato, Kuroda, King, Ando QPT, Valla, TCI Story}, scanning tunneling spectroscopies (STM) \cite{Roushan, Zhang_STM, Alpichshev, Cheng, Hanaguri, Jungpil, Haim Nature physics BiSe, Okada}, transport \cite{Qu, Analytis, Peng, Hadar1, Chen1, Ando_BTS, FuChun, Yayu, Samarth, Morpurgo, Fuhrer, Checkelsky}, optical methods \cite{CD1, Hsieh_SHG, Hancock, CD2, Ultrafast, Ultrafast2, Light, Kerr, Basov} and even on nano-structured samples \cite{Xue Nature physics QL, Peng, Hadar1}. Here, we will review the procedures for i) separating intrinsic bulk bands from surface electronic structures using incident photon energy modulated ARPES, ii) mapping the surface electronic structure across the Kramers momenta to establish the topologically non-trivial nature of the surface states, iii) using spin-ARPES to map the spin texture of the surface states to reveal topological quantum numbers and Berry's phases and iv) measuring the topological parent compounds to establish the microscopic origins of 3D topological order. These will be discussed in the context of Bi$_{1-x}$Sb$_x$, which was the first 3D topological insulator to be experimentally discovered in nature and a textbook example of how this method is applied. The confluence of three factors, having a detailed spectroscopic procedure to measure 3D topological order, their discovery in fairly simple bulk semiconductors and being able to work at room temperatures, has led to worldwide efforts to study 3D topological physics and led to over 100 compounds being identified as 3D topological insulators to date.

\begin{figure*}
\includegraphics[scale=0.62,clip=true, viewport=0.0in 0in 10.8in 7.0in]{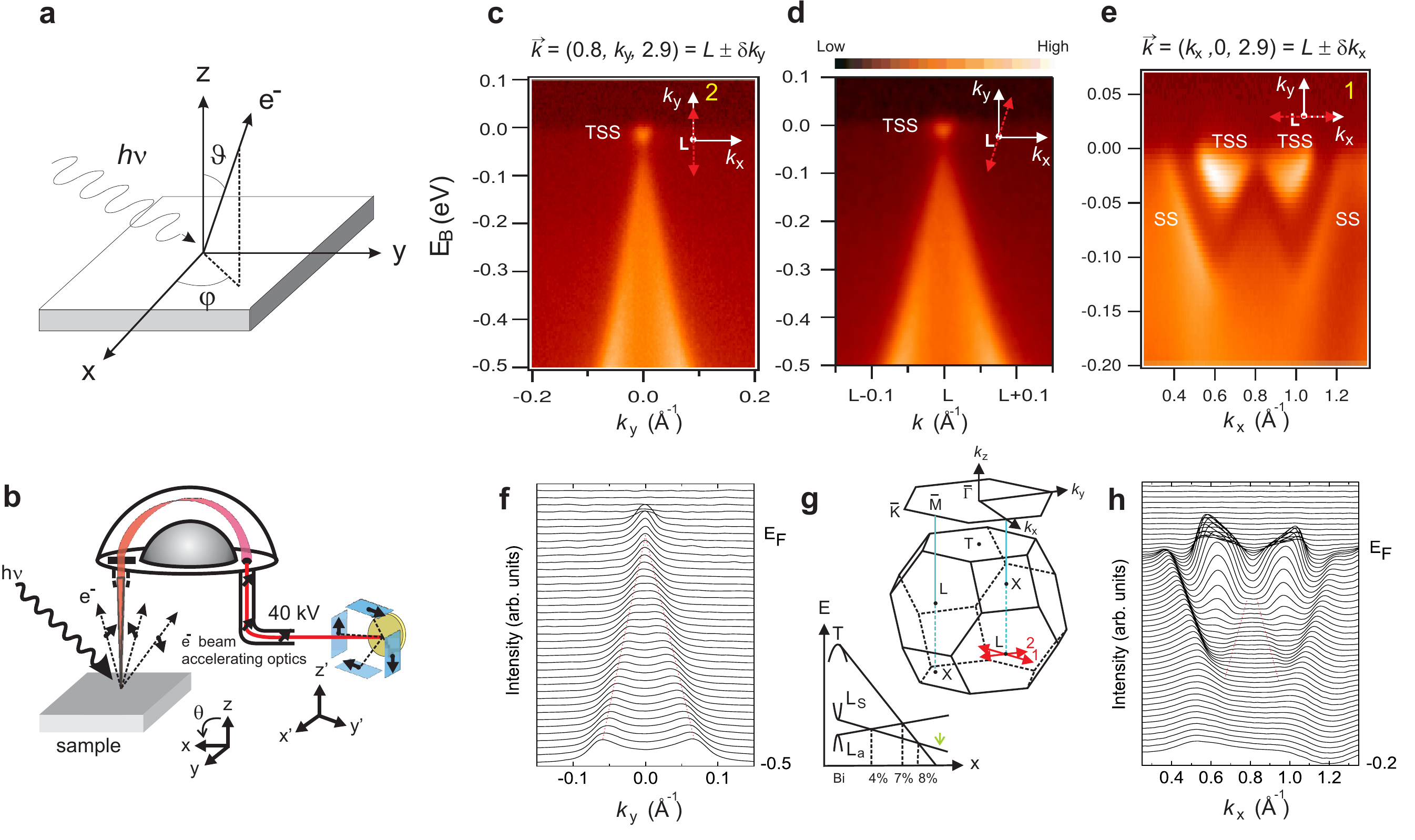}
\caption{\label{fig:BiSb_Fig1} \textbf{The first 3D topological insulator (2007) with in-gap Fermi level: Dirac-like dispersion signalling band inversion via spin-orbit interaction} \textbf{a}, Schematic of an ARPES experimental geometry. The kinetic energy of photoelectrons and their angles of emission ($\theta$,$\phi$) determine its electronic structure. \textbf{b}, Energy and momentum analysis take place through a hemispherical analyzer and the spin analysis is performed using a Mott detector. Selected ARPES intensity
maps of Bi$_{0.9}$Sb$_{0.1}$ are shown along three $\vec{k}$-space
cuts through the L point of the bulk 3D Brillouin zone (BZ). The
presented data are taken in the third BZ with L$_z$ = 2.9 \AA$^{-1}$
with a photon energy of 29 eV. The cuts are along \textbf{c}, the
$k_y$ direction, \textbf{d}, a direction rotated by approximately
$10^{\circ}$ from the $k_y$ direction, and \textbf{e}, the $k_x$
direction. Each cut shows a $\Lambda$-shaped bulk band whose tip
lies below the Fermi level signalling a bulk gap. The (topological) surface states
are denoted (T)SS and are all identified in Fig.\ref{fig:BiSb_Fig2}. \textbf{f}, Momentum distribution curves (MDCs)
corresponding to the intensity map in \textbf{c}. \textbf{h}, Log
scale plot of the MDCs corresponding to the intensity map in
\textbf{e}. The red lines are guides to the eye for the bulk
features in the MDCs. \textbf{g}, Schematic of the bulk 3D BZ of
Bi$_{1-x}$Sb$_x$ and the 2D BZ of the projected (111) surface. The
high symmetry points $\bar{\Gamma}$, $\bar{M}$ and $\bar{K}$ of the
surface BZ are labeled. Schematic evolution of bulk band energies as
a function of $x$ is shown. The L band inversion transition occurs
at $x \approx 0.04$, where a 3D gapless Dirac point is realized, and
the composition we study here (for which $x = 0.1$) is indicated by
the green arrow. A more detailed phase diagram based on our
experiments is shown in Fig.\ref{fig:BiSb_Fig3}c. Previous works on these compounds did not probe the topological aspects. [Adapted from D. Hsieh $et$ $al.$, \textit{Nature} \textbf{452}, 970 (2008) [Completed and submitted in \textbf{2007}] \cite{10}].}
\end{figure*}

\section{Separation of insulating bulk from metallic surface states using incident photon energy modulated ARPES}

Three-dimensional topological order is predicted to occur in semiconductors with an inverted band gap, therefore 3D topological insulators are often searched for in systems where a band gap inversion is known to take place as a function of some control parameter. The experimental signature of being in the vicinity of a bulk band inversion is that the bulk band dispersion should be described by the massive Dirac equation rather than Schr/"odinger equation, since the system must be described by the massless Dirac equation exactly at the bulk band inversion point.

\begin{figure*}
\includegraphics[scale=0.65,clip=true, viewport=0.0in 0in 10.0in 7.4in]{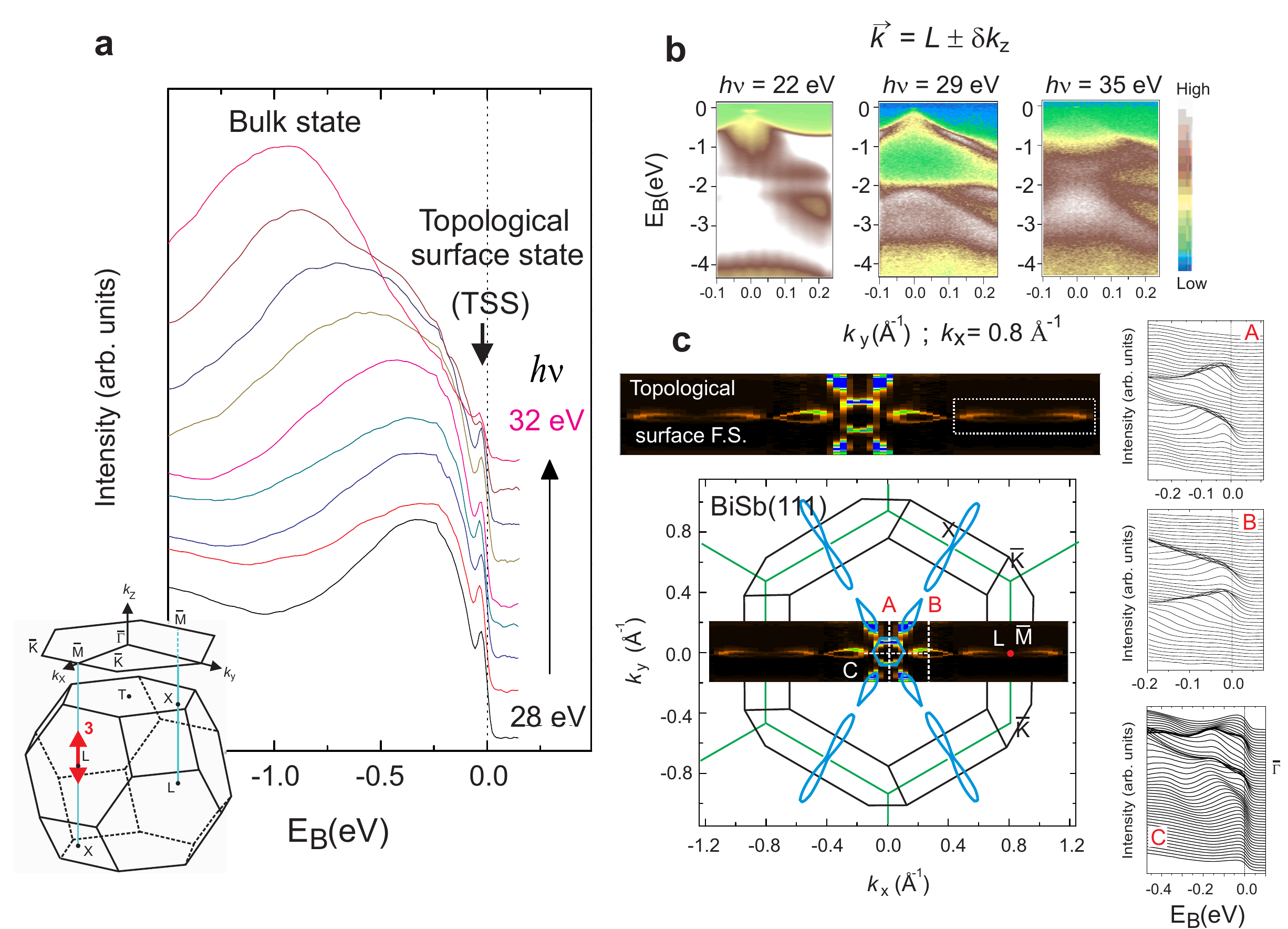}
\caption{\label{fig:BiSb_Fig2} \textbf{The first 3D topological insulator  with in-gap Fermi level (2007): Topological Surface States and electronic band dispersion along the $\mathbf{k_z}$-direction in momentum space.} Surface states are experimentally
identified by studying their out-of-plane momentum dispersion
through the systematic variation of incident photon energy.
\textbf{a}, Energy distribution curves (EDCs) of
Bi$_{0.9}$Sb$_{0.1}$ with electrons at the Fermi level ($E_\textrm{F}$)
maintained at a fixed in-plane momentum of ($k_x$=0.8 \AA$^{-1}$,
$k_y$=0.0 \AA$^{-1}$) are obtained as a function of incident photon
energy to identify states that exhibit no dispersion perpendicular
to the (111)-plane along the direction shown by the double-headed
arrow labeled ``3" in the inset. Selected EDC data sets with photon energies of 28 eV to
32 eV in steps of 0.5 eV are shown for clarity. The non-energy
dispersive ($k_z$ independent) peaks near $E_\textrm{F}$ are the topological surface
states (TSS). \textbf{b}, ARPES intensity maps along cuts parallel to
$k_y$ taken with electrons at $E_\textrm{F}$ fixed at $k_x$ = 0.8 \AA$^{-1}$
with respective photon energies of $h \nu$ = 22 eV, 29 eV and 35 eV. The faint $\Lambda$-shaped band at $h \nu$ = 22 eV and
$h \nu$ = 35 eV shows some overlap with the bulk valence band at L
($h \nu$ = 29 eV), suggesting that it is a resonant surface state
degenerate with the bulk state in some limited k-range near $E_\textrm{F}$.
The flat band of intensity centered about $-$2 eV in the $h \nu$ =
22 eV scan originates from Bi 5\textit{d} core level emission from second
order light. \textbf{c}, Projection of the bulk BZ (black lines)
onto the (111) surface BZ (green lines). Overlay (enlarged in inset)
shows the high resolution Fermi surface (FS) of the metallic SS
mode, which was obtained by integrating the ARPES intensity (taken
with $h \nu$ = 20 eV) from $-$15 meV to 10 meV relative to $E_\textrm{F}$.
The six tear-drop shaped lobes of the surface FS close to
$\bar{\Gamma}$ (center of BZ) show some intensity variation between
them that is due to the relative orientation between the axes of the
lobes and the axis of the detector slit. The six-fold symmetry was
however confirmed by rotating the sample in the $k_x-k_y$ plane.
EDCs corresponding to the cuts A, B and C are also shown; these
confirm the gapless character of the surface states in bulk
insulating Bi$_{0.9}$Sb$_{0.1}$. [Adapted from D. Hsieh $et$ $al.$, \textit{Nature} \textbf{452}, 970 (2008) \cite{10}].}
\end{figure*}

The early theoretical treatments \cite{11,Murakami} focused on the
strongly spin-orbit coupled, band-inverted Bi$_{1-x}$Sb$_x$ series
as a possible realization of 3D topological order for the following reason. Bismuth is a semimetal with strong spin-orbit interactions. Its band
structure is believed to feature an indirect negative gap between
the valence band maximum at the T point of the bulk Brillouin zone
(BZ) and the conduction band minima at three equivalent L points
\cite{Lenoir,Liu} (here we generally refer to these as a single
point, L). The valence and conduction bands at L are derived from
antisymmetric (L$_a$) and symmetric (L$_s$) $p$-type orbitals,
respectively, and the effective low-energy Hamiltonian at this point
is described by the (3+1)-dimensional relativistic Dirac equation
\cite{Wolff, Fukuyama, Buot}. The resulting dispersion relation,
$E(\vec{k}) = \pm \sqrt{ {(\vec{v})}^2 {(\vec{k})}^2 + \Delta^2}
\approx \vec{v} \cdot \vec{k}$, is highly linear owing to the
combination of an unusually large band velocity $\vec{v}$ and a
small gap $\Delta$ (such that $\lvert \Delta / \lvert \vec{v} \rvert
\rvert \approx 5 \times 10^{-3} $\AA$^{-1}$) and has been used to
explain various peculiar properties of bismuth \cite{Wolff,
Fukuyama, Buot}. Substituting bismuth with antimony is believed to
change the critical energies of the band structure as follows (see
Fig.\ref{fig:BiSb_Fig1}). At an Sb concentration of $x \approx 4\%$, the gap $\Delta$
between L$_a$ and L$_s$ closes and a massless three-dimensional (3D)
Dirac point is realized. As $x$ is further increased this gap
re-opens with inverted symmetry ordering, which leads to a change in
sign of $\Delta$ at each of the three equivalent L points in the BZ.
For concentrations greater than $x \approx 7\%$ there is no overlap
between the valence band at T and the conduction band at L, and the
material becomes an inverted-band insulator. Once the band at T
drops below the valence band at L, at $x \approx 8\%$, the system
evolves into a direct-gap insulator whose low energy physics is
dominated by the spin-orbit coupled Dirac particles at L
\cite{11,Lenoir}.

High-momentum-resolution angle-resolved photoemission spectroscopy (Fig.\ref{fig:BiSb_Fig1}a \& b)
performed with varying incident photon energy (IPEM-ARPES) allows
for measurement of electronic band dispersion along various momentum
space ($\vec{k}$-space) trajectories in the 3D bulk BZ. ARPES
spectra taken along two orthogonal cuts through the L point of the
bulk BZ of Bi$_{0.9}$Sb$_{0.1}$ are shown in Figs \ref{fig:BiSb_Fig1} c and e. A
$\Lambda$-shaped dispersion whose tip lies less than 50 meV below
the Fermi energy ($E_\textrm{F}$) can be seen along both directions.
Additional features originating from surface states that do not
disperse with incident photon energy are also seen. Owing to the
finite intensity between the bulk and surface states, the exact
binding energy ($E_\textrm{B}$) where the tip of the $\Lambda$-shaped band
dispersion lies is unresolved. The linearity of the bulk
$\Lambda$-shaped bands is observed by locating the peak positions at
higher $E_\textrm{B}$ in the momentum distribution curves (MDCs), and the
energy at which these peaks merge is obtained by extrapolating
linear fits to the MDCs. Therefore 50 meV represents a lower bound
on the energy gap $\Delta$ between L$_a$ and L$_s$. The magnitude of
the extracted band velocities along the $k_x$ and $k_y$ directions
are $7.9 \pm 0.5 \times 10^4$ ms$^{-1}$ and $10.0 \pm 0.5 \times
10^5$ ms$^{-1}$, respectively, which are similar to the tight
binding values $7.6 \times 10^4$ ms$^{-1}$ and $9.1 \times 10^5$
ms$^{-1}$ calculated for the L$_a$ band of bismuth \cite{Liu}. Our
data are consistent with the extremely small effective mass of
$0.002m_e$ (where $m_e$ is the electron mass) observed in
magneto-reflection measurements on samples with $x = 11\%$
\cite{Hebel}. The Dirac point in graphene, co-incidentally, has a
band velocity ($|v_F| \approx 10^6$ ms$^{-1}$) \cite{Zhang}
comparable to what we observe for Bi$_{0.9}$Sb$_{0.1}$, but its
spin-orbit coupling is several orders of magnitude weaker, and the only known method of inducing a gap
in the Dirac spectrum of graphene is by coupling to an external
chemical substrate (which is a trivial gap) \cite{Zhou}. The Bi$_{1-x}$Sb$_x$ series thus
provides a rare opportunity to study relativistic Dirac Hamiltonian
physics in a 3D condensed matter system where the intrinsic (rest)
mass gap can be easily tuned.

\begin{figure*}
\includegraphics[scale=0.65,clip=true, viewport=2.5in 0in 16.5in 8.7in]{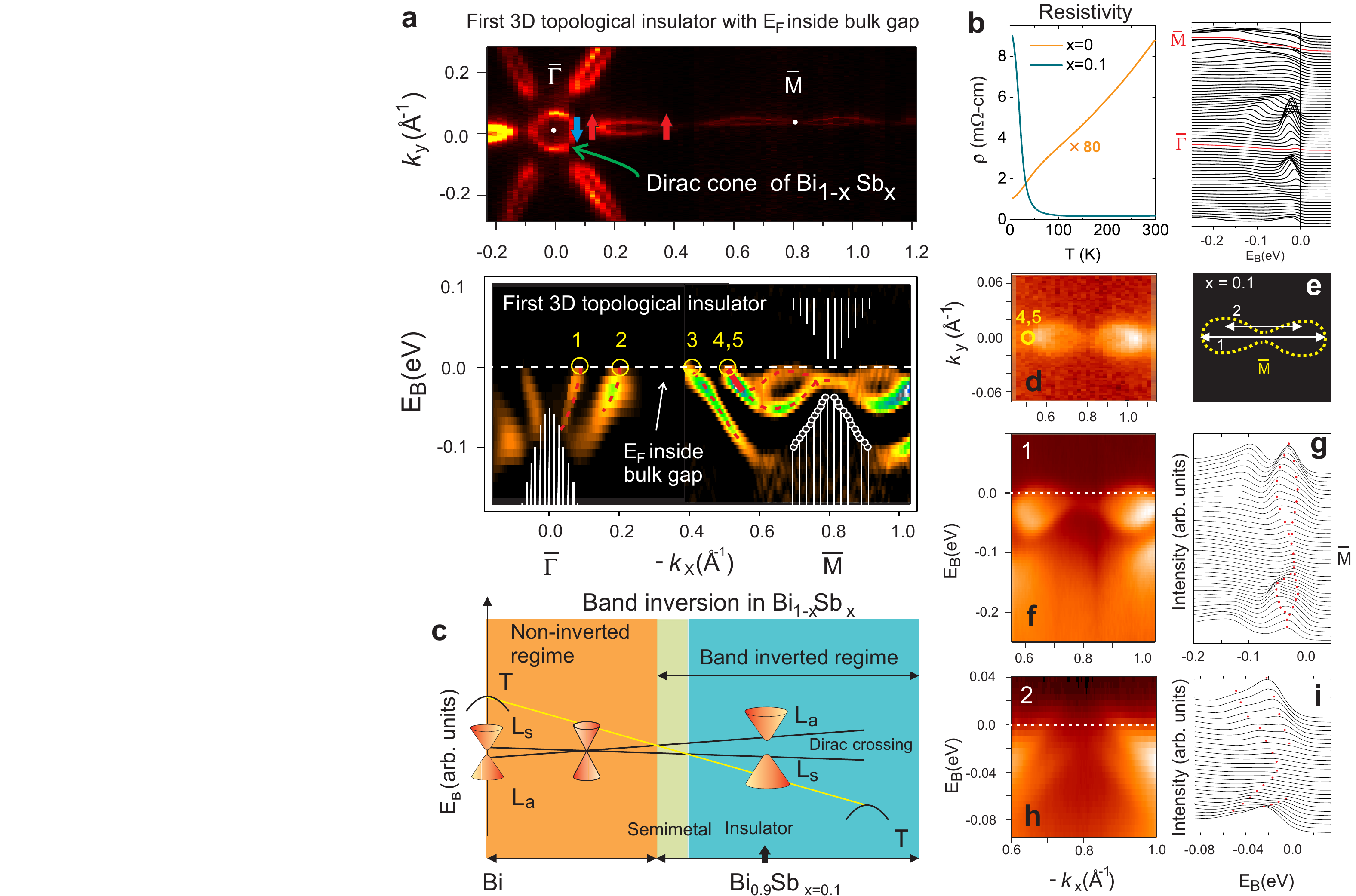}
\caption{\label{fig:BiSb_Fig3} \textbf{The first 3D topological insulator with in-gap Fermi level: The topological gapless
surface states in bulk insulating (observed via the bulk band dispersion gap) Bi$_{0.9}$Sb$_{0.1}$.} \textbf{a},
The surface Fermi surface and surface band dispersion second derivative image (SDI) of
Bi$_{0.9}$Sb$_{0.1}$ along $\bar{\Gamma} - \bar{M}$. The shaded
white area shows the projection of the bulk bands based on ARPES
data, as well as a rigid shift of the tight binding bands to sketch
the unoccupied bands above the Fermi level. To maintain high
momentum resolution, data were collected in two segments of momentum
space, then the intensities were normalized using background level
above the Fermi level. A non-intrinsic flat band of intensity near
$E_\textrm{F}$ generated by the SDI analysis was rejected to isolate the
intrinsic dispersion. The Fermi crossings of the surface state are
denoted by yellow circles, with the band near $-k_x \approx 0.5$
\AA$^{-1}$ counted twice owing to double degeneracy. The red lines
are guides to the eye. An in-plane rotation of the sample by
$60^{\circ}$ produced the same surface state dispersion. The EDCs
along $\bar{\Gamma} - \bar{M}$ are shown to the right. There are a
total of five crossings from $\bar{\Gamma} - \bar{M}$ which
indicates that these surface states are topologically non-trivial.
The number of surface state crossings in a material (with an odd
number of Dirac points) is related to the topological Z$_2$
invariant (see text). \textbf{b}, The resistivity curves of Bi and
Bi$_{0.9}$Sb$_{0.1}$ reflect the contrasting transport behaviours.
The presented resistivity curve for pure bismuth has been multiplied
by a factor of 80 for clarity. \textbf{c}, Schematic variation of
bulk band energies of Bi$_{1-x}$Sb$_x$ as a function of $x$ (based
on band calculations and on \cite{11, Lenoir}).
Bi$_{0.9}$Sb$_{0.1}$ is a direct gap bulk Dirac point insulator well
inside the inverted-band regime, and its surface forms a
``topological metal'' - the 2D analogue of the 1D edge states in
quantum spin Hall systems. \textbf{d}, ARPES intensity integrated
within $\pm 10$ meV of $E_\textrm{F}$ originating solely from the surface
state crossings. The image was plotted by stacking along the
negative $k_x$ direction a series of scans taken parallel to the
$k_y$ direction. \textbf{e}, Outline of Bi$_{0.9}$Sb$_{0.1}$ surface
state ARPES intensity near $E_\textrm{F}$ measured in \textbf{d}. White lines
show scan directions ``1'' and ``2''. \textbf{f}, Surface band
dispersion along direction ``1'' taken with $h \nu$ = 28 eV and the
corresponding EDCs (\textbf{g}). The surface Kramers degenerate
point, critical in determining the topological Z$_2$ class of a band
insulator, is clearly seen at $\bar{M}$, approximately $15 \pm 5$
meV below $E_\textrm{F}$. (We note that the scans are taken along the
negative $k_x$ direction, away from the bulk L point.) \textbf{h},
Surface band dispersion along direction ``2'' taken with $h \nu$
 = 28 eV and the corresponding EDCs (\textbf{i}). This scan no longer
passes through the $\bar{M}$-point, and the observation of two well
separated bands indicates the absence of Kramers degeneracy as
expected, which cross-checks the result in (\textbf{a}). [Adapted from D. Hsieh $et$ $al.$, \textit{Nature} \textbf{452}, 970 (2008)\cite{10}].}
\end{figure*}

Studying the band dispersion perpendicular to the sample surface
provides a way to differentiate bulk states from surface states in a
3D material. To visualize the near-$E_\textrm{F}$ dispersion along the 3D L-X
cut (X is a point that is displaced from L by a $k_z$ distance of
3$\pi/c$, where $c$ is the lattice constant), in Fig.\ref{fig:BiSb_Fig2}a we plot
energy distribution curves (EDCs), taken such that electrons at
$E_\textrm{F}$ have fixed in-plane momentum $(k_x, k_y)$ = (L$_x$, L$_y$) =
(0.8 \AA$^{-1}$, 0.0 \AA$^{-1}$), as a function of photon energy
($h\nu$). There are three prominent features in the EDCs: a
non-dispersing, $k_z$ independent, peak centered just below $E_\textrm{F}$ at
about $-$0.02 eV; a broad non-dispersing hump centered near $-$0.3
eV; and a strongly dispersing hump that coincides with the latter
near $h\nu$ = 29 eV. To understand which bands these features
originate from, we show ARPES intensity maps along an in-plane cut
$\bar{K} \bar{M} \bar{K}$ (parallel to the $k_y$ direction) taken
using $h\nu$ values of 22 eV, 29 eV and 35 eV, which correspond to
approximate $k_z$ values of L$_z -$ 0.3 \AA$^{-1}$, L$_z$, and L$_z$
+ 0.3 \AA$^{-1}$ respectively (Fig.\ref{fig:BiSb_Fig2}b). At $h\nu$ = 29 eV, the low
energy ARPES spectral weight reveals a clear $\Lambda$-shaped band
close to $E_\textrm{F}$. As the photon energy is either increased or
decreased from 29 eV, this intensity shifts to higher binding
energies as the spectral weight evolves from the $\Lambda$-shaped
into a $\cup$-shaped band. Therefore the dispersive peak in Fig.2a
comes from the bulk valence band, and for $h\nu$ = 29 eV the high
symmetry point L = (0.8, 0, 2.9) appears in the third bulk BZ. In
the maps of Fig.\ref{fig:BiSb_Fig2}b with respective $h\nu$ values of 22 eV and 35 eV,
overall weak features near $E_\textrm{F}$ that vary in intensity remain even
as the bulk valence band moves far below $E_\textrm{F}$. The survival of
these weak features over a large photon energy range (17 to 55 eV)
supports their surface origin. The non-dispersing feature centered
near $-0.3$ eV in Fig.\ref{fig:BiSb_Fig2}a comes from the higher binding energy
(valence band) part of the full spectrum of surface states, and the
weak non-dispersing peak at $-0.02$ eV reflects the low energy part
of the surface states that cross $E_\textrm{F}$ away from the $\bar{M}$ point
and forms the surface Fermi surface (Fig.\ref{fig:BiSb_Fig2}c).

\begin{figure*}
\includegraphics[scale=0.56,clip=true, viewport=0.0in 0.5in 11.3in 6.5in]{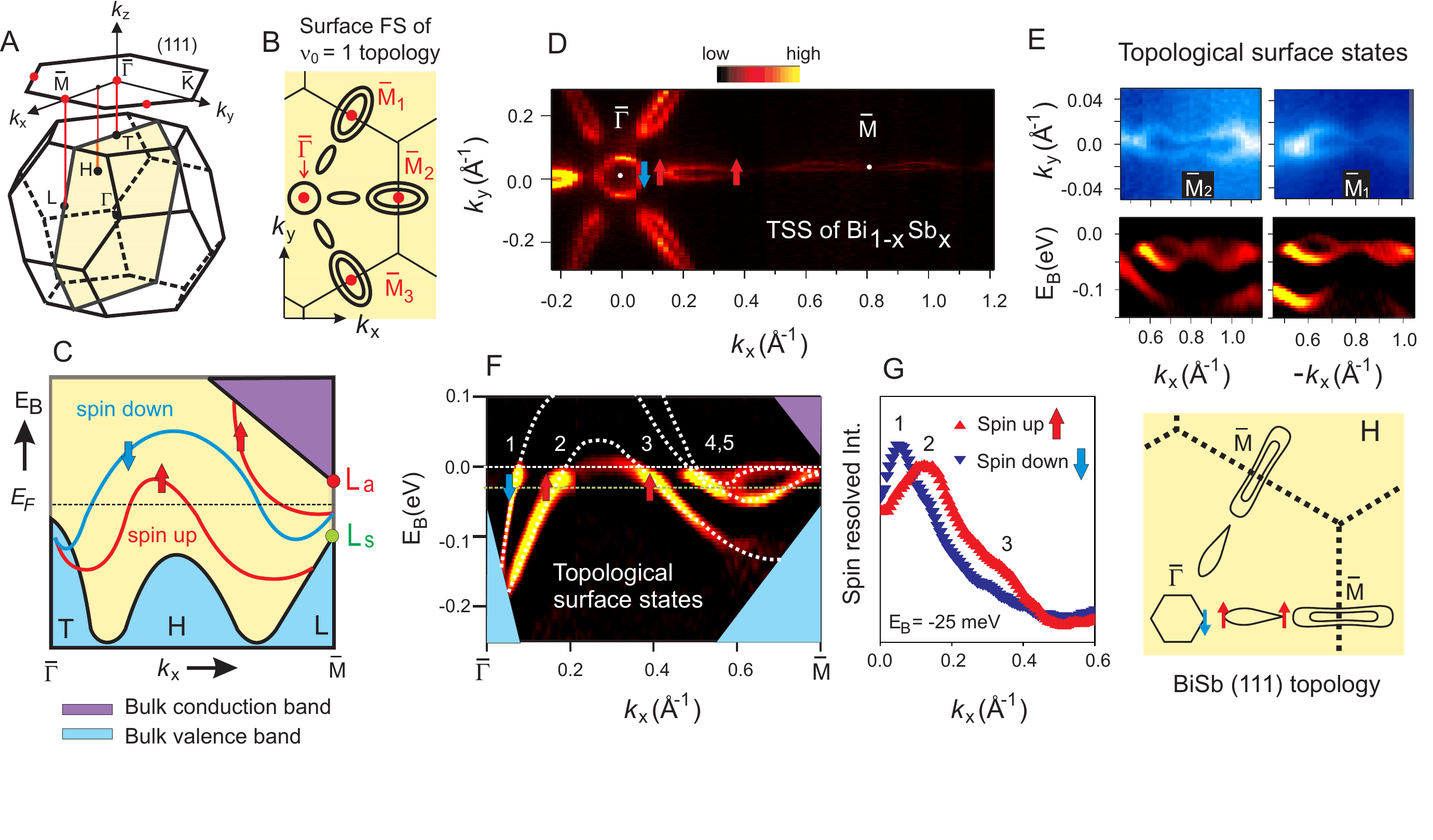}
\caption{\label{Sb_Fig1} \textbf{Spin texture of a topological insulator encodes Z$_2$ topological order of the bulk} (A) Schematic
sketches of the bulk Brillouin zone (BZ) and (111) surface BZ of the
Bi$_{1-x}$Sb$_x$ crystal series. The high symmetry points
(L,H,T,$\Gamma$,$\bar{\Gamma}$,$\bar{M}$,$\bar{K}$) are identified. (B)
Schematic of Fermi surface pockets formed by the surface states (SS)
of a topological insulator that carries a Berry's phase. (C) Partner
switching band structure topology: Schematic of spin-polarized SS
dispersion and connectivity between $\bar{\Gamma}$ and $\bar{M}$
required to realize the FS pockets shown in panel-(B). $L_a$ and
$L_s$ label bulk states at $L$ that are antisymmetric and symmetric
respectively under a parity transformation (see text). (D)
Spin-integrated ARPES intensity map of the SS of
Bi$_{0.91}$Sb$_{0.09}$ at $E_\textrm{F}$. Arrows point in the measured
direction of the spin. (E) High resolution ARPES intensity map of
the SS at $E_\textrm{F}$ that enclose the $\bar{M}_1$ and $\bar{M}_2$ points.
Corresponding band dispersion (second derivative images) are shown
below. The left right asymmetry of the band dispersions are due to
the slight offset of the alignment from the
$\bar{\Gamma}$-$\bar{M}_1$($\bar{M}_2$) direction. (F) Surface band
dispersion image along the $\bar{\Gamma}$-$\bar{M}$ direction showing
five Fermi level crossings. The intensity of bands 4,5 is scaled up
for clarity (the dashed white lines are guides to the eye). The
schematic projection of the bulk valence and conduction bands are
shown in shaded blue and purple areas. (G) Spin-resolved momentum
distribution curves presented at $E_\textrm{B}$ = $-$25 meV showing single
spin degeneracy of bands at 1, 2 and 3. Spin up and down correspond
to spin pointing along the +$\hat{y}$ and -$\hat{y}$ direction
respectively. (H) Schematic of the spin-polarized surface FS
observed in our experiments. It is consistent with a $\nu_0$ = 1
topology (compare (B) and (H)). [Adapted from D. Hsieh $et$ $al.$, \textit{Science} \textbf{323}, 919 (2009)\cite{Science}].}
\end{figure*}

\section{Winding number count: Counting of surface Fermi surfaces enclosing Kramers points to identify topologically non-trivial surface spin-textured states}

\begin{figure*}
\includegraphics[width=13cm]{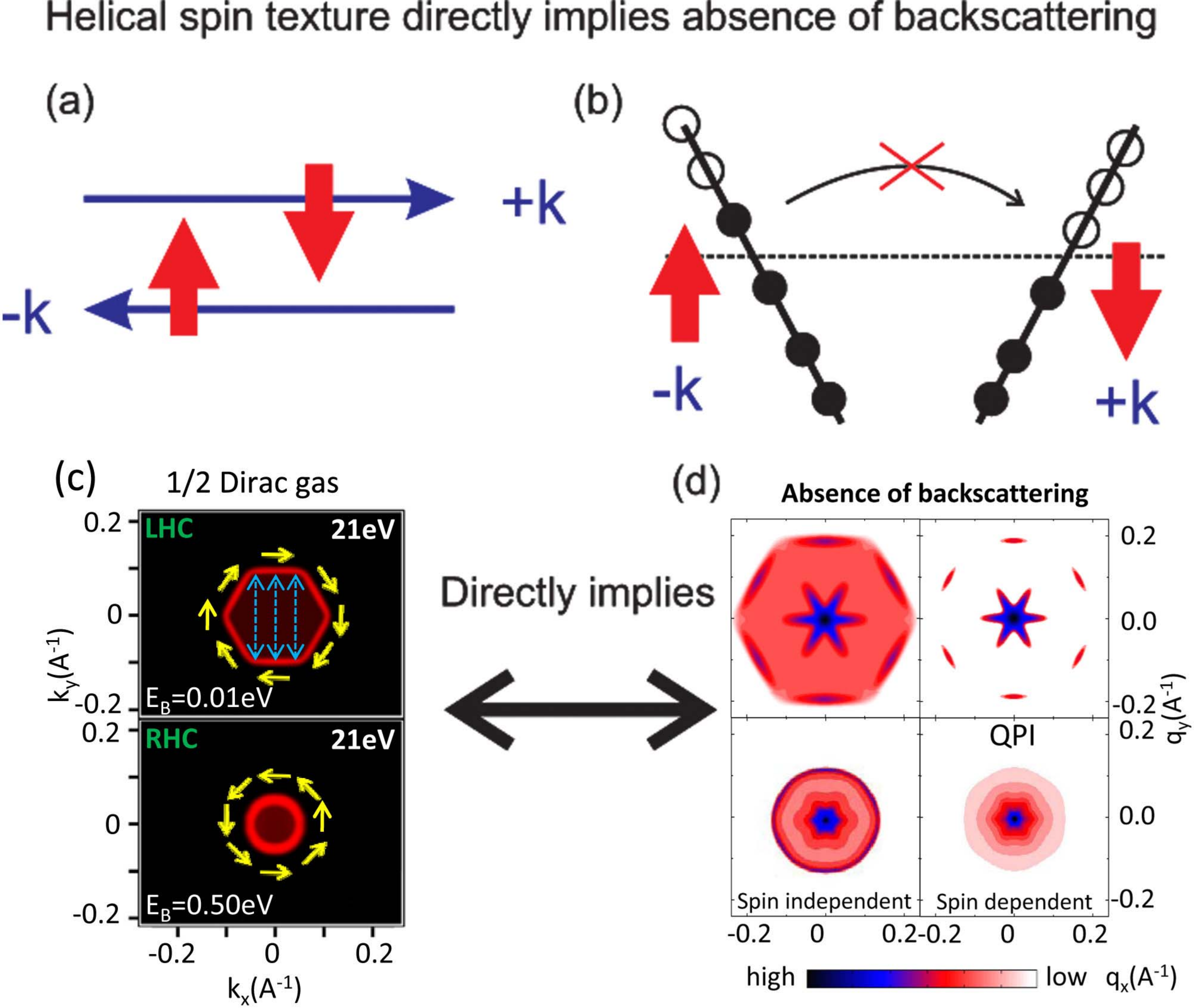}
\caption{\label{backscatter} \textbf{Helical spin texture naturally leads to absence of elastic backscattering for surface transport: No "U" turn on a 3D topological insulator surface.} (a) Our measurement of a helical spin texture in both Bi$_{1-x}$Sb$_x$ and in Bi$_2$Se$_3$ directly shows that there is (b) an absence of backscattering. (c) ARPES measured FSs are shown with spin
directions based on polarization measurements. L(R)HC stands for left(right)-handed chirality. (d) Spin independent and spin dependent scattering profiles on FSs in (c) relevant for surface quasi-particle transport are shown which is sampled by the quasi-particle interference (QPI) modes. [Adapted from S.-Y. Xu $et$ $al.$, \textit{Science} \textbf{332} 560 (2011). \cite{Xu}]}
\end{figure*}

Having established the existence of an energy gap in the bulk state
of Bi$_{0.9}$Sb$_{0.1}$ (Figs \ref{fig:BiSb_Fig1} and \ref{fig:BiSb_Fig2}) and observed linearly
dispersive bulk bands uniquely consistent with strong spin-orbit
coupling model calculations \cite{Wolff, Fukuyama, Buot, Liu}, we now discuss the topological character of its
surface states, which are found to be gapless (Fig.\ref{fig:BiSb_Fig2}c). In general,
the states at the surface of spin-orbit coupled compounds are
allowed to be spin split owing to the loss of space inversion
symmetry $[E(k,\uparrow) = E(-k,\uparrow)]$. However, as required by
Kramers' theorem, this splitting must go to zero at the four time
reversal invariant momenta (TRIM) in the 2D surface BZ. As discussed
in \cite{11, Fu:STI2}, along a path connecting two TRIM in the
same BZ, the Fermi energy inside the bulk gap will intersect these
singly degenerate surface states either an even or odd number of
times. When there are an even number of surface state crossings, the
surface states are topologically Z$_2$ trivial because disorder or correlations can remove \emph{pairs}
of such crossings by pushing the surface bands entirely above or
below $E_\textrm{F}$. When there are an odd number of crossings, however, at
least one surface state must remain gapless, which makes it
non-trivial \cite{11, Murakami, Fu:STI2}. The existence of such
topologically non-trivial surface states can be theoretically
predicted on the basis of the \emph{bulk} band structure only, using
the Z$_2$ invariant that is related to the quantum Hall Chern number
\cite{14}. Materials with band structures with Z$_2 = +1$  ($\nu_0$ = 0)
are ordinary Bloch band insulators that are topologically equivalent
to the filled shell atomic insulator, and are predicted to exhibit
an even number (including zero) of surface state crossings.
Materials with bulk band structures with Z$_2 = -1$ ($\nu_0$ = 1) on the other
hand, which are expected to exist in rare systems with strong
spin-orbit coupling acting as an internal quantizing magnetic field
on the electron system \cite{Haldane(P-anomaly)}, and inverted bands
at an odd number of high symmetry points in their bulk 3D BZs, are
predicted to exhibit an odd number of surface state crossings,
precluding their adiabatic continuation to the atomic insulator
\cite{11, Murakami, Fu:STI2, 15,
8, 7}. Such ``topological quantum Hall metals''
\cite{Fu:STI2, 15} cannot be realized in a purely 2D
electron gas system such as the one realized at the interface of
GaAs/GaAlAs systems.

\begin{figure}
\includegraphics[scale=0.5,clip=true, viewport=0.0in 1.7in 6.8in 7.8in]{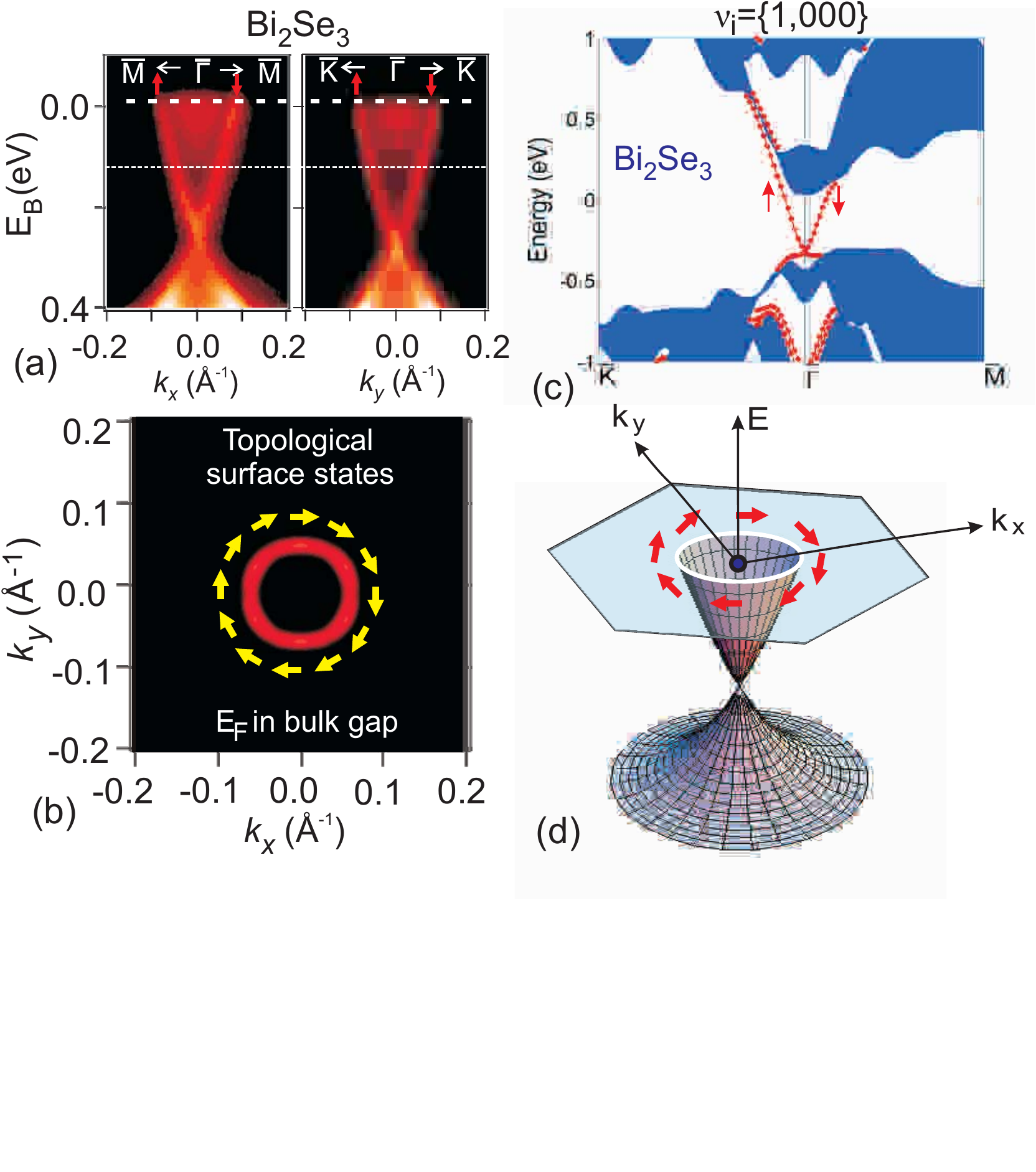}
\caption{\label{RMP_Fig12} \textbf{Spin-momentum locked helical fermions reveal topological-order of the bulk:} Spin-momentum locked helical
surface Dirac fermions are hallmark signatures of topological
insulators. (a) ARPES data for Bi$_2$Se$_3$ reveals surface electronic
states with a single spin-polarized Dirac cone. The
surface Fermi surface (b) exhibits a chiral left-handed spin
texture. Data is from a sample doped to a Fermi level [thin line in (a)] that is in the bulk gap. ARPES measurements are carried out in P-polarizations mode which couples strongly to the (dominant) p$_z$ orbital component of the surface state wavefunction. (c) Surface electronic structure of Bi$_2$Se$_3$ computed
in the local density approximation. The shaded regions describe
bulk states, and the red lines are surface states. (d)
Schematic of the spin polarized surface state dispersion in
the Bi$_2$X$_3$ topological insulators. [Adapted from Y. Xia $et$ $al$., \textit{Nature Phys.} \textbf{5}, 398 (2009).\cite{Xia}].}
\end{figure}

\begin{figure*}
\includegraphics[scale=0.7,clip=true, viewport=-0.0in 0in 11.0in 5.5in]{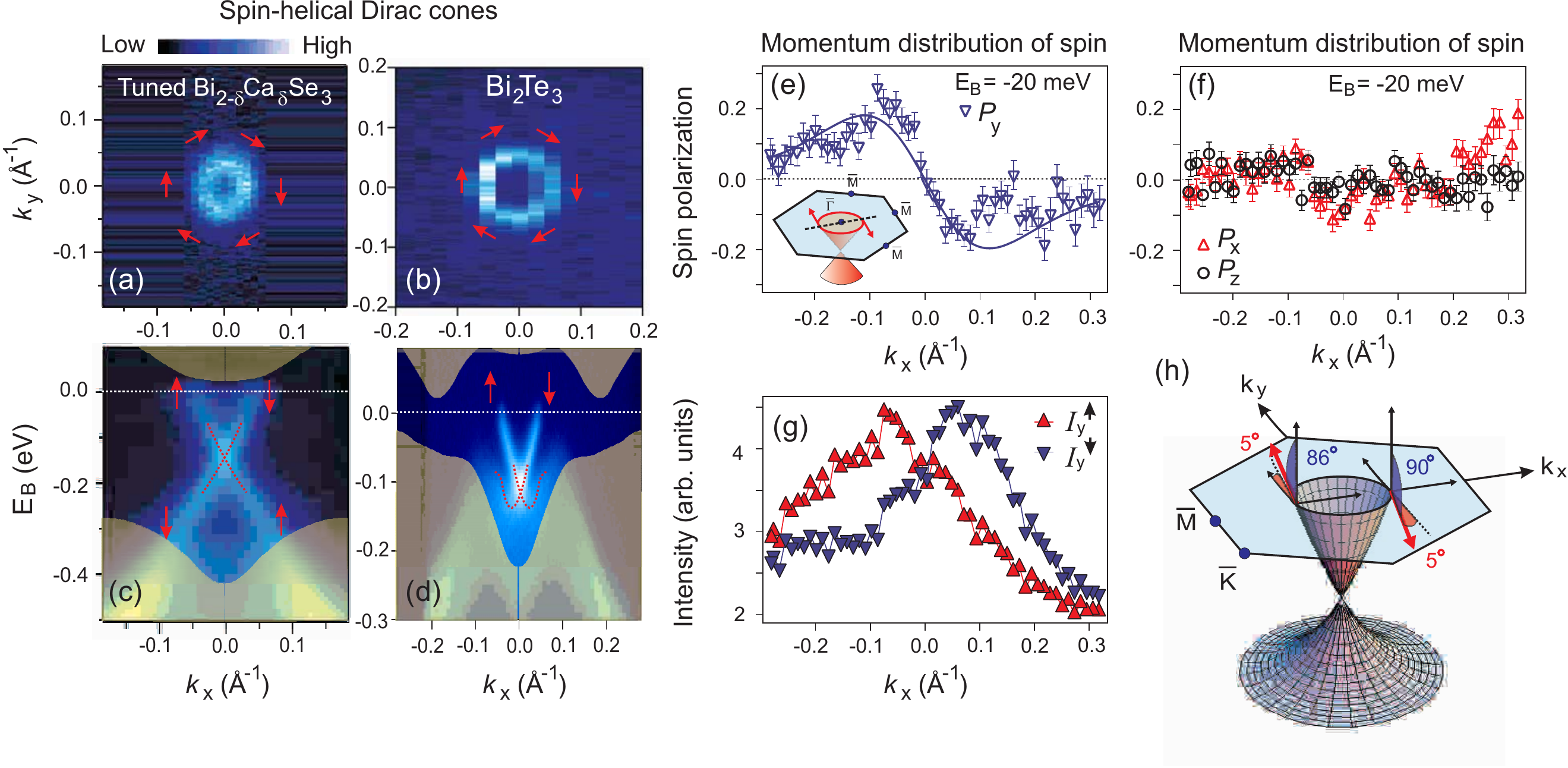}
\caption{\label{fig:Fig1} \textbf{First detection of Topological-Order: spin-momentum locking of spin-helical Dirac electrons in Bi$_2$Se$_3$ and Bi$_2$Te$_3$ using spin-resolved ARPES.} (a) ARPES intensity map at $E_\textrm{F}$ of the (111)
surface of tuned Bi$_{2-\delta}$Ca$_{\delta}$Se$_3$ (see text) and
(b) the (111) surface of Bi$_2$Te$_3$. Red arrows denote the
direction of spin around the Fermi surface. (c) ARPES dispersion of
tuned Bi$_{2-\delta}$Ca$_{\delta}$Se$_3$ and (d) Bi$_2$Te$_3$ along
the $k_x$ cut. The dotted red lines are guides to the eye. The
shaded regions in (c) and (d) are our calculated projections of the
bulk bands of pure Bi$_2$Se$_3$ and Bi$_2$Te$_3$ onto the (111)
surface respectively. (e) Measured $y$ component of
spin-polarization along the $\bar{\Gamma}$-$\bar{M}$ direction at $E_\textrm{B}$
= -20 meV, which only cuts through the surface states. Inset shows a
schematic of the cut direction. (f) Measured $x$ (red triangles) and
$z$ (black circles) components of spin-polarization along the
$\bar{\Gamma}$-$\bar{M}$ direction at $E_\textrm{B}$ = -20 meV. Error bars in (e)
and (f) denote the standard deviation of $P_{x,y,z}$, where typical
detector counts reach $5\times10^5$; Solid lines are numerical fits
\cite{21}. (g) Spin-resolved spectra obtained from the $y$ component
spin polarization data. The non-Lorentzian lineshape of the
$I_y^{\uparrow}$ and $I_y^{\downarrow}$ curves and their non-exact
merger at large $|k_{x}|$ is due to the time evolution of the
surface band dispersion, which is the dominant source of statistical
uncertainty. a.u., arbitrary units. (h) Fitted values of the spin
polarization vector \textbf{P} = ($S_x$,$S_y$,$S_z$) are
(sin(90$^{\circ}$)cos(-95$^{\circ}$),
sin(90$^{\circ}$)sin(-95$^{\circ}$), cos(90$^{\circ}$)) for
electrons with +$k_x$ and (sin(86$^{\circ}$)cos(85$^{\circ}$),
sin(86$^{\circ}$)sin(85$^{\circ}$), cos(86$^{\circ}$)) for electrons
with -$k_x$, which demonstrates the topological helicity of the
spin-Dirac cone. ARPES measurements are carried out in P-polarizations mode which couples strongly to the (dominant) p$_z$ orbital component of the surface state wavefunction. The angular uncertainties are of order
$\pm$10$^{\circ}$ and the magnitude uncertainty is of order
$\pm$0.15. [Adapted from D. Hsieh $et$ $al.$, \textit{Nature} \textbf{460}, 1101 (2009). \cite{Nature_2009}].}
\end{figure*}

In our experimental case, namely the (111) surface of
Bi$_{0.9}$Sb$_{0.1}$, the four TRIM are located at $\bar{\Gamma}$
and three $\bar{M}$ points that are rotated by $60^{\circ}$ relative
to one another. Owing to the three-fold crystal symmetry (A7 bulk
structure) and the observed mirror symmetry of the surface Fermi
surface across $k_x = 0$ (Fig.\ref{fig:BiSb_Fig2}), these three $\bar{M}$ points are
equivalent (and we henceforth refer to them as a single point,
$\bar{M}$). The mirror symmetry $[E(k_y) = E(-k_y)]$ is also
expected in this system. The complete details of the surface state dispersion observed in our
experiments along a path connecting $\bar{\Gamma}$ and $\bar{M}$ are
shown in Fig.\ref{fig:BiSb_Fig3}a; finding this information is made possible by our
experimental separation of surface states from bulk states. As for
bismuth (Bi), two surface bands emerge from the bulk band continuum
near $\bar{\Gamma}$ to form a central electron pocket and an
adjacent hole lobe \cite{Ast:Bi1, Hochst,Hofmann}. It has been
established that these two bands result from the spin-splitting of a
surface state and are thus singly degenerate \cite{Hirahara,
Hofmann}. On the other hand, the surface band that crosses $E_\textrm{F}$ at
$-k_x \approx 0.5$ \AA$^{-1}$, and forms the narrow electron pocket
around $\bar{M}$, is clearly doubly degenerate, as far as we can
determine within our experimental resolution. This is indicated by
its splitting below $E_\textrm{F}$ between $-k_x \approx 0.55$ \AA$^{-1}$ and
$\bar{M}$, as well as the fact that this splitting goes to zero at
$\bar{M}$ in accordance with Kramers theorem. In semimetallic single
crystal bismuth, only a single surface band is observed to form the
electron pocket around $\bar{M}$ \cite{Hengsberger, Ast:Bi2}.
Moreover, this surface state overlaps, hence becomes degenerate
with, the bulk conduction band at L (L projects to the surface
$\bar{M}$ point) owing to the semimetallic character of Bi (Fig.\ref{fig:BiSb_Fig3}b).
In Bi$_{0.9}$Sb$_{0.1}$ on the other hand, the states near $\bar{M}$
fall completely inside the bulk energy gap preserving their purely
surface character at $\bar{M}$ (Fig.\ref{fig:BiSb_Fig3}a). The surface Kramers doublet
point can thus be defined in the bulk insulator (unlike in Bi
\cite{Hirahara,Ast:Bi1, Hochst, Hofmann, Hengsberger, Ast:Bi2}) and
is experimentally located in Bi$_{0.9}$Sb$_{0.1}$ samples to lie
approximately 15 $\pm$ 5 meV below $E_\textrm{F}$ at $\vec{k} = \bar{M}$
(Fig.\ref{fig:BiSb_Fig3}a). For the precise location of this Kramers point, it is
important to demonstrate that our alignment is strictly along the
$\bar{\Gamma} - \bar{M}$ line. To do so, we contrast high resolution
ARPES measurements taken along the $\bar{\Gamma} - \bar{M}$ line
with those that are slightly offset from it (Fig.\ref{fig:BiSb_Fig3}e). Figs \ref{fig:BiSb_Fig3}f-i show
that with $k_y$ offset from the Kramers point at $\bar{M}$ by less
than 0.02 \AA$^{-1}$, the degeneracy is lifted and only one band
crosses $E_\textrm{F}$ to form part of the bow-shaped electron distribution
(Fig.\ref{fig:BiSb_Fig3}d). Our finding of five surface state crossings (an odd rather
than an even number) between $\bar{\Gamma}$ and $\bar{M}$ (Fig.\ref{fig:BiSb_Fig3}a),
confirmed by our observation of the Kramers degenerate point at the
TRIM, indicates that these gapless surface states are topologically
non-trivial. This corroborates our bulk electronic structure result
that Bi$_{0.9}$Sb$_{0.1}$ is in the insulating band-inverted (Z$_2 =
-1$ ($\nu_0$ = 1)) regime (Fig.\ref{fig:BiSb_Fig3}c), which contains an odd number of bulk (gapped)
Dirac points.

\begin{figure*}
\includegraphics[scale=0.6,clip=true, viewport=0.0in 0.7in 11.0in 5.4in]{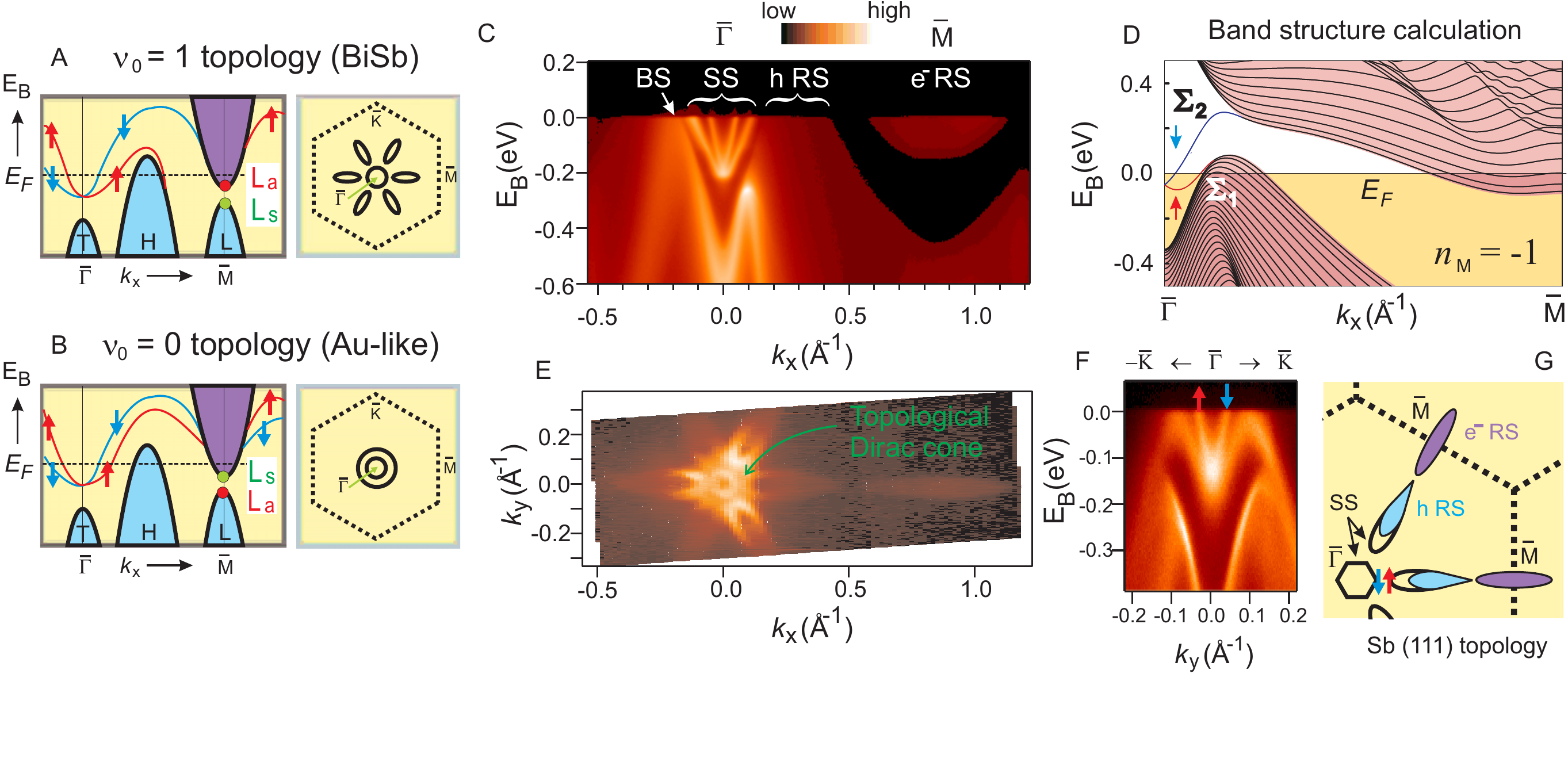}
\caption{\label{Sb_Fig2} \textbf{Topological character of parent compound revealed on the (111) surface states.} Schematic of the bulk band structure (shaded
areas) and surface band structure (red and blue lines) of Sb near
$E_\textrm{F}$ for a (A) topologically non-trivial and (B) topological
trivial (gold-like) case, together with their corresponding surface
Fermi surfaces are shown. (C) Spin-integrated ARPES spectrum of
Sb(111) along the $\bar{\Gamma}$-$\bar{M}$ direction. The surface states
are denoted by SS, bulk states by BS, and the hole-like resonance
states and electron-like resonance states by h RS and e$^-$ RS
respectively. (D) Calculated surface state band structure of Sb(111)
based on the methods in \cite{20,Liu}. The continuum bulk energy bands are
represented with pink shaded regions, and the lines show the
discrete bands of a 100 layer slab. The red and blue single bands,
denoted $\Sigma_1$ and $\Sigma_2$, are the surface states bands with
spin polarization $\langle \vec{P} \rangle \propto +\hat{y}$ and
$\langle \vec{P} \rangle \propto -\hat{y}$ respectively. (E) ARPES
intensity map of Sb(111) at $E_\textrm{F}$ in the $k_x$-$k_y$ plane. The only
one FS encircling $\bar{\Gamma}$ seen in the data is formed by the
inner V-shaped SS band seen in panel-(C) and (F). The outer V-shaped
band bends back towards the bulk band best seen in data in
panel-(F). (F) ARPES spectrum of Sb(111) along the
$\bar{\Gamma}-\bar{K}$ direction shows that the outer V-shaped SS band
merges with the bulk band. (G) Schematic of the surface FS of
Sb(111) showing the pockets formed by the surface states (unfilled)
and the resonant states (blue and purple). The purely surface state
Fermi pocket encloses only one Kramers degenerate point
($\vec{k}_T$), namely, $\bar{\Gamma}$(=$\vec{k}_T$), therefore
consistent with the $\nu_0$ = 1 topological classification of Sb
which is different from Au (compare (B) and (G)). As discussed in
the text, the hRS and e$^-$RS count trivially. [Adapted from D. Hsieh $et$ $al.$, \textit{Science} \textbf{323}, 919 (2009) \cite{Science}].}
\end{figure*}

\begin{figure*}
\includegraphics[scale=0.62,clip=true, viewport=0.0in 0in 11.0in 6.0in]{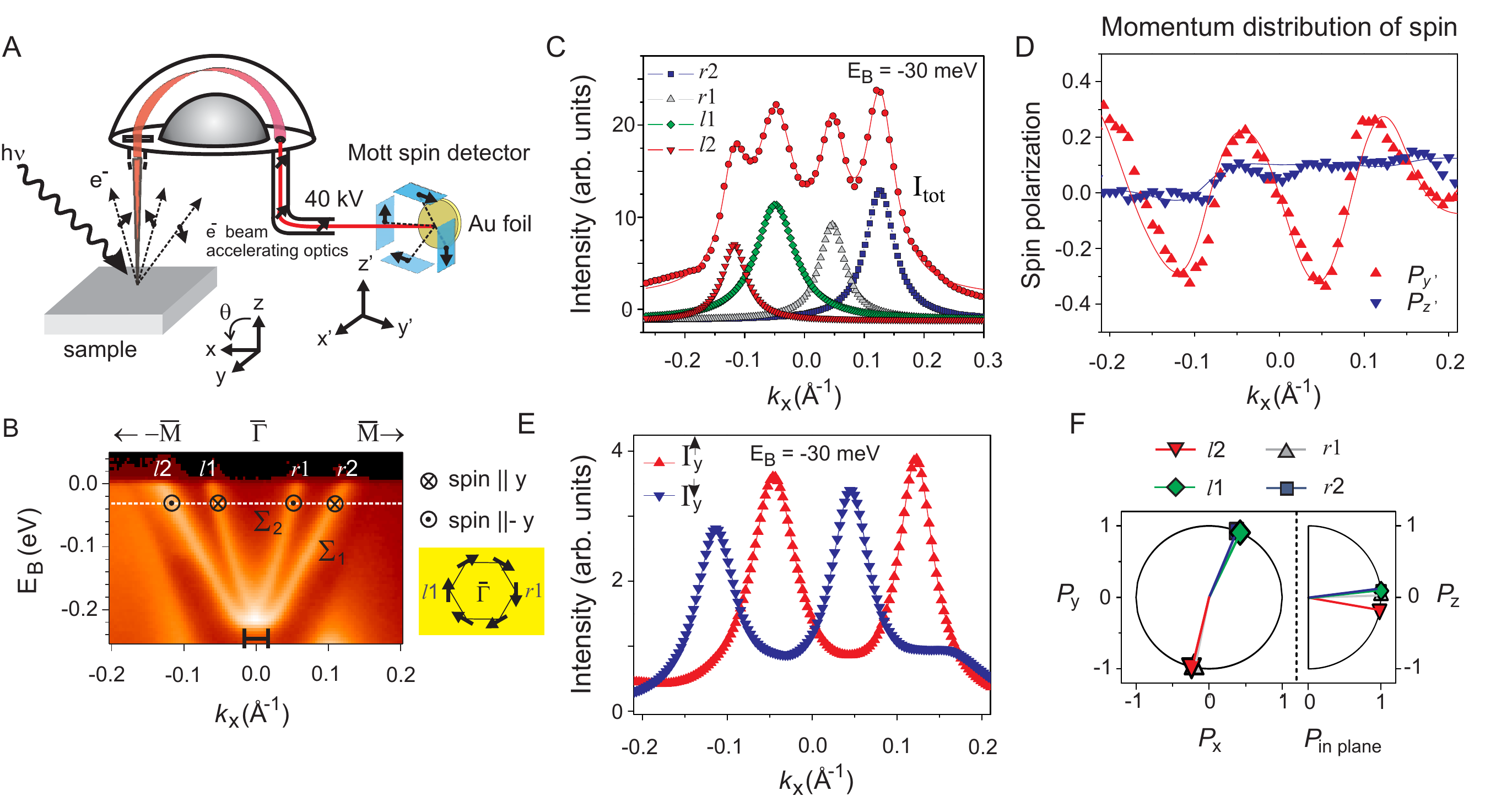}
\caption{\label{Sb_Fig3} \textbf{Spin-texture of topological surface states and topological chirality.} (A) Experimental geometry of the spin-resolved ARPES
study. At normal emission ($\theta$ = 0$^{\circ}$), the sensitive
$y'$-axis of the Mott detector is rotated by 45$^{\circ}$ from the
sample $\bar{\Gamma}$ to $-\bar{M}$ ($\parallel -\hat{x}$) direction,
and the sensitive $z'$-axis of the Mott detector is parallel to the
sample normal ($\parallel \hat{z}$). (B) Spin-integrated ARPES
spectrum of Sb(111) along the $-\bar{M}$-$\bar{\Gamma}$-$\bar{M}$
direction. The momentum splitting between the band minima is
indicated by the black bar and is approximately 0.03 \AA$^{-1}$. A
schematic of the spin chirality of the central FS based on the
spin-resolved ARPES results is shown on the right. (C) Momentum
distribution curve of the spin averaged spectrum at $E_\textrm{B}$ = $-$30
meV (shown in (B) by white line), together with the Lorentzian peaks
of the fit. (D) Measured spin polarization curves (symbols) for the
detector $y'$ and $z'$ components together with the fitted lines
using the two-step fitting routine \cite{26}. (E) Spin-resolved spectra
for the sample $y$ component based on the fitted spin polarization
curves shown in (D). Up (down) triangles represent a spin direction
along the +(-)$\hat{y}$ direction. (F) The in-plane and out-of-plane
spin polarization components in the sample coordinate frame obtained
from the spin polarization fit. Overall spin-resolved data and the
fact that the surface band that forms the central electron pocket
has $\langle \vec{P} \rangle \propto -\hat{y}$ along the +$k_x$
direction, as in (E), suggest a left-handed chirality (schematic in
(B) and see text for details). [Adapted from D. Hsieh $et$ $al.$, \textit{Science} \textbf{323}, 919 (2009) \cite{Science}].}
\end{figure*}

These experimental results taken collectively strongly suggest that
Bi$_{0.9}$Sb$_{0.1}$ is quite distinct from graphene \cite{Zhang,
Novoselov} and represents a novel state of quantum matter: a
strongly spin-orbit coupled insulator with an odd number of Dirac
points with a negative Z$_2$ topological Hall phase, which realizes
the ``parity anomaly without Fermion doubling". These works further
demonstrates a general methodology for possible future
investigations of \emph{novel topological orders} in exotic quantum
matter.

\section{Spin-resolving the surface states to identify the non-trivial topological phase and establish a 2D helical metal protected from backscattering}

\begin{figure}
\includegraphics[width=8.5cm]{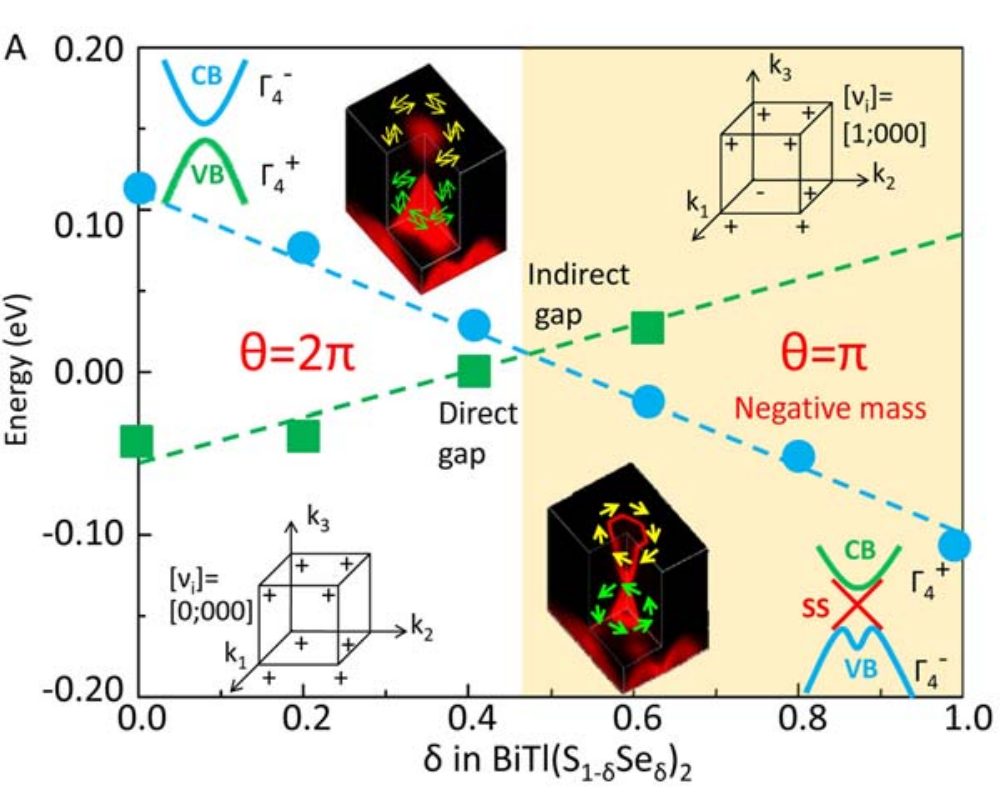}
\caption{\label{QPT} \textbf{Observation of bulk band inversion across a topological quantum phase transition}. Energy levels of $\Gamma_4^-$. (blue circles) and $\Gamma_4^+$ (green squares) bands are obtained from ARPES measurements as a function of composition $\delta$. CB: conduction band; VB: valence
band. Parity eigenvalues (+ or ) of Bloch states are
shown. The topological invariants, $\nu_i$, obtained from the
parity eigenvalues are presented as [$\theta$/$\pi$ = $\nu_0$;$\nu_1$,$\nu_2$,$\nu_3$] where $\theta$ =
$\pi \nu_0$ is the axion angle \cite{12} and $\nu_0$ is the strong invariant. ARPES measurements are carried out in P-polarizations mode which couples strongly to the (dominant) p$_z$ orbital component of the surface state wavefunction. [Adapted from S.-Y. Xu $et$ $al.$, \textit{Science} \textbf{332} 560 (2011). \cite{Xu}]}
\end{figure}

Strong topological materials are distinguished from ordinary
materials such as gold by a topological quantum number, $\nu_0$ = 1
or 0 respectively \cite{14,15}. For Bi$_{1-x}$Sb$_x$, theory has shown
that $\nu_0$ is determined solely by the character of the bulk
electronic wave functions at the $L$ point in the three-dimensional
(3D) Brillouin zone (BZ). When the lowest energy conduction band
state is composed of an antisymmetric combination of atomic $p$-type
orbitals ($L_a$) and the highest energy valence band state is
composed of a symmetric combination ($L_s$), then $\nu_0$ = 1, and
vice versa for $\nu_0$ = 0 \cite{11}. Although the bonding nature
(parity) of the states at $L$ is not revealed in a measurement of
the bulk band structure, the value of $\nu_0$ can be determined from
the spin-textures of the surface bands that form when the bulk is
terminated. In particular, a $\nu_0$ = 1 topology requires the
terminated surface to have a Fermi surface (FS) that supports a
non-zero Berry's phase (odd as opposed to even multiple of $\pi$),
which is not realizable in an ordinary spin-orbit material.

In a general inversion symmetric spin-orbit insulator, the bulk
states are spin degenerate because of a combination of space
inversion symmetry $[E(\vec{k},\uparrow) = E(-\vec{k},\uparrow)]$
and time reversal symmetry $[E(\vec{k},\uparrow) =
E(-\vec{k},\downarrow)]$. Because space inversion symmetry is broken
at the terminated surface, the spin degeneracy of surface bands can
be lifted by the spin-orbit interaction [19-21]. However, according
to Kramers theorem [16], they must remain spin degenerate at four
special time reversal invariant momenta ($\vec{k}_T$ =
$\bar{\Gamma}$, $\bar{M}$) in the surface BZ [11], which for the (111)
surface of Bi$_{1-x}$Sb$_x$ are located at $\bar{\Gamma}$ and three
equivalent $\bar{M}$ points [see Fig.\ref{Sb_Fig1}(A)].

Depending on whether $\nu_0$ equals 0 or 1, the Fermi surface
pockets formed by the surface bands will enclose the four
$\vec{k}_T$ an even or odd number of times respectively. If a Fermi
surface pocket does not enclose $\vec{k}_T$ (= $\bar{\Gamma}$,
$\bar{M}$), it is irrelevant for the Z$_2$ topology \cite{11,20}. Because the wave
function of a single electron spin acquires a geometric phase factor
of $\pi$ \cite{16} as it evolves by 360$^{\circ}$ in momentum space along
a Fermi contour enclosing a $\vec{k}_T$, an odd number of Fermi
pockets enclosing $\vec{k}_T$ in total implies a $\pi$ geometrical
(Berry's) phase \cite{11}. In order to realize a $\pi$ Berry's phase the
surface bands must be spin-polarized and exhibit a partner switching
\cite{11} dispersion behavior between a pair of $\vec{k}_T$. This means
that any pair of spin-polarized surface bands that are degenerate at
$\bar{\Gamma}$ must not re-connect at $\bar{M}$, or must separately
connect to the bulk valence and conduction band in between
$\bar{\Gamma}$ and $\bar{M}$. The partner switching behavior is realized
in Fig. \ref{Sb_Fig1}(C) because the spin down band connects to and is
degenerate with different spin up bands at $\bar{\Gamma}$ and $\bar{M}$.

We first investigate the spin properties of the topological
insulator phase \cite{Science}. Spin-integrated ARPES \cite{19} intensity maps of the
(111) surface states of insulating Bi$_{1-x}$Sb$_x$ taken at the
Fermi level ($E_\textrm{F}$) [Figs \ref{Sb_Fig1}(D)\&(E)] show that a hexagonal FS
encloses $\bar{\Gamma}$, while dumbbell shaped FS pockets that are
much weaker in intensity enclose $\bar{M}$. By examining the surface
band dispersion below the Fermi level [Fig.\ref{Sb_Fig1}(F)] it is clear that
the central hexagonal FS is formed by a single band (Fermi crossing
1) whereas the dumbbell shaped FSs are formed by the merger of two
bands (Fermi crossings 4 and 5) \cite{10}.

\begin{figure}
\includegraphics[scale=1,clip=true, viewport=0.0in 0.0in 5.4in 3.5in]{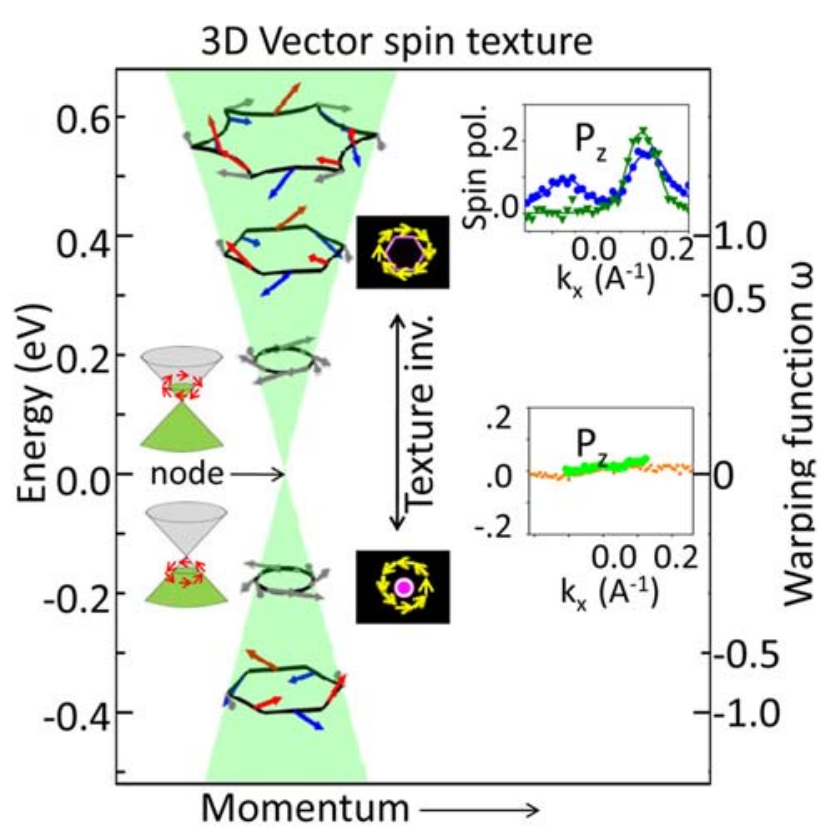}
\caption{\label{texture} \textbf{Spin texture evolution of topological surface bands as a function of energy away from the Dirac node} (left axis) and geometrical warping factor $\omega$ (right axis). The warping factor is defined as $\omega = \frac{k_F(\bar{\Gamma}-\bar{M})-k_F(\bar{\Gamma}-\bar{K})}{k_F(\bar{\Gamma}-\bar{M})+k_F(\bar{\Gamma}-\bar{K})}\times\frac{2+\sqrt{3}}{2-\sqrt{3}}$ where $\omega$ = 0, $\omega$ = 1, and $\omega$ $>$ 1 implies circular, hexagonal and snowflake-shaped FSs respectively. The sign of $\omega$ indicates texture chirality for LHC (+) or RHC (-). Insets: out-of-plane 3D spin-polarization measurements at corresponding FSs. ARPES measurements are carried out in P-polarizations mode which couples strongly to the (dominant) p$_z$ orbital component of the surface state wavefunction. [Adapted from S.-Y. Xu $et$ $al.$, \textit{Science} \textbf{332} 560 (2011). \cite{Xu}]}
\end{figure}

This band dispersion resembles the partner switching dispersion
behavior characteristic of topological insulators. To check this
scenario and determine the topological index $\nu_0$, we have
carried out spin-resolved photoemission spectroscopy. Fig.\ref{Sb_Fig1}(G) shows
a spin-resolved momentum distribution curve taken along the
$\bar{\Gamma}$-$\bar{M}$ direction at a binding energy $E_\textrm{B}$ = $-$25 meV
[Fig.\ref{Sb_Fig1}(G)]. The data reveal a clear difference between the spin-up
and spin-down intensities of bands 1, 2 and 3, and show that bands 1
and 2 have opposite spin whereas bands 2 and 3 have the same spin
(detailed analysis discussed later in text). The former observation
confirms that bands 1 and 2 form a spin-orbit split pair, and the
latter observation suggests that bands 2 and 3 (as opposed to bands
1 and 3) are connected above the Fermi level and form one band. This
is further confirmed by directly imaging the bands through raising
the chemical potential via doping. Irrelevance of bands 2 and 3 to the topology is
consistent with the fact that the Fermi surface pocket they form
does not enclose any $\vec{k}_T$. Because of a dramatic intrinsic
weakening of signal intensity near crossings 4 and 5, and the small
energy and momentum splitting of bands 4 and 5 lying at the
resolution limit of modern spin-resolved ARPES spectrometers, no
conclusive spin information about these two bands can be drawn from
the methods employed in obtaining the data sets in Figs \ref{Sb_Fig1}(G)\&(H).
However, whether bands 4 and 5 are both singly or doubly degenerate
does not change the fact that an odd number of spin-polarized FSs
enclose the $\vec{k}_T$, which provides evidence that
Bi$_{1-x}$Sb$_x$ has $\nu_0$ = 1 and that its surface supports a
non-trivial Berry's phase. This directly implies an absence of backscattering in electronic transport along the surface (Fig.\ref{backscatter}), which has been re-confirmed by numerous scanning tunneling microscopy studies that show quasi-particle interference patterns that can only be modeled assuming an absence of backscattering \cite{Roushan,Alpichshev,Zhang_STM}.

Shortly after the discovery of Bi$_{1-x}$Sb$_x$, physicists sought to find a simpler version of a 3D topological insulator consisting of a single surface state instead of five. This is because the surface structure of Bi$_{1-x}$Sb$_x$ was rather complicated and the band gap was rather small. This motivated
a search for topological insulators with a larger band gap
and simpler surface spectrum. A second generation of
3D topological insulator materials, especially
Bi$_2$Se$_3$, offers the potential for topologically protected
behavior in ordinary crystals at room temperature
and zero magnetic field. Starting in 2008, work by the Princeton
group used spin-ARPES and first-principles calculations to
study the surface band structure of Bi$_2$Se$_3$ and observe
the characteristic signature of a topological insulator in
the form of a single Dirac cone that is spin-polarized (Figs \ref{RMP_Fig12} and \ref{fig:Fig1}) such that it also carries a non-trivial Berry's phase \cite{Xia,Nature_2009}. Concurrent theoretical work by \cite{Zhang_nphys} used electronic structure methods to show
that Bi$_2$Se$_3$ is just one of several new large band-gap
topological insulators. These other materials were soon after also identified using this ARPES technique we describe \cite{Chen,Hsieh_PRL}.

\begin{figure}
\includegraphics[scale=0.32,clip=true, viewport=0.0in 0in 11.0in 5.0in]{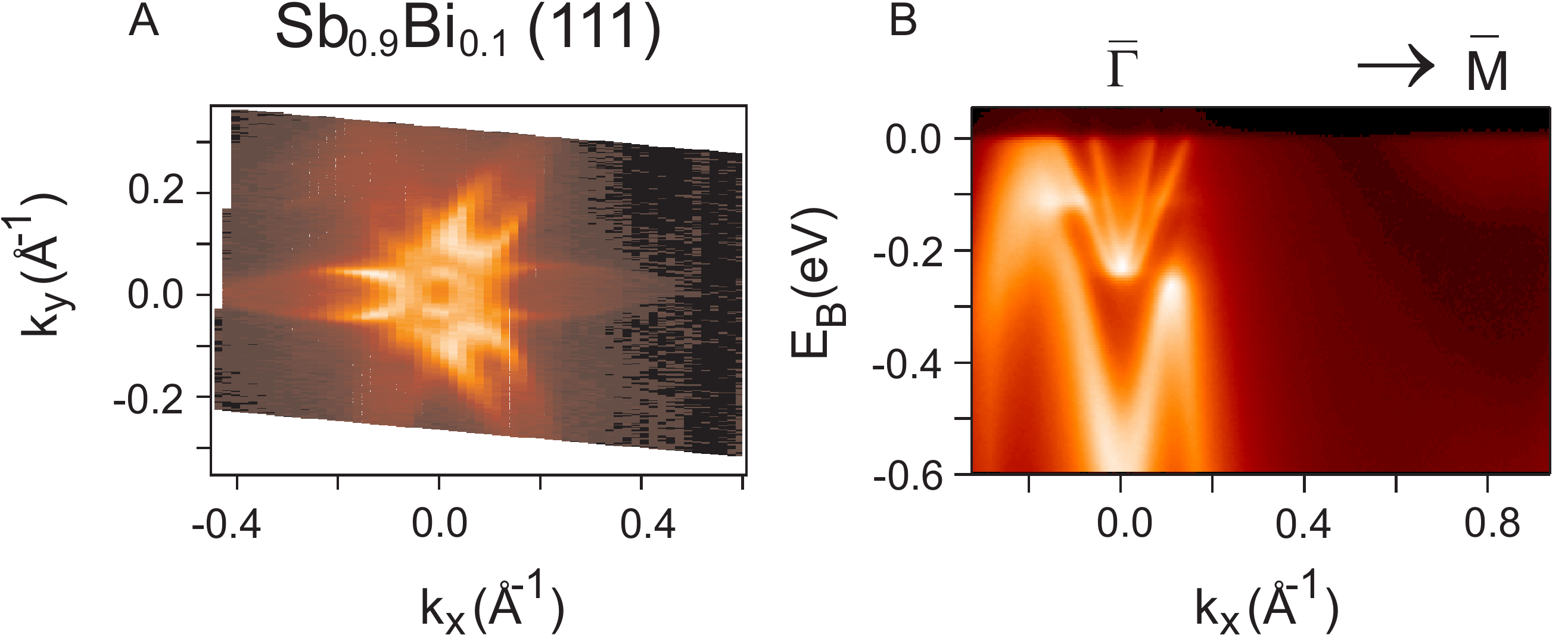}
\caption{\label{fig:Sb_FigS5} \textbf{Robustness against disorder} Spin split surface states survive alloying
disorder in Sb$_{0.9}$Bi$_{0.1}$. (\textbf{A}) ARPES intensity map
at $E_\textrm{F}$ of single crystal Sb$_{0.9}$Bi$_{0.1}$(111) in the
$k_x$-$k_y$ plane taken using 20 eV photons. (\textbf{B}) ARPES
intensity map of Sb$_{0.9}$Bi$_{0.1}$(111) along the
$\bar{\Gamma}$-$\bar{M}$ direction taken with $h\nu$ = 22 eV photons.
The band dispersion is not symmetric about $\bar{\Gamma}$ because of
the three-fold rotational symmetry of the bulk states about the
$\langle111\rangle$ axis. [Adapted from D. Hsieh $et$ $al.$, \textit{Science} \textbf{323}, 919 (2009) \cite{Science}].}
\end{figure}

The Bi$_2$Se$_3$ surface state is found from spin-ARPES and theory to be a nearly idealized single Dirac
cone as seen from the experimental data in Figs.\ref{RMP_Fig12} and \ref{RMP_Fig13}. An added advantage is that Bi$_2$Se$_3$ is stoichiometric (i.e., a pure
compound rather than an alloy such as Bi$_{1-x}$Sb$_x$) and
hence can be prepared, in principle, at higher purity.
While the topological insulator phase is predicted to be
quite robust to disorder, many experimental probes of
the phase, including ARPES of the surface band structure,
are clearer in high-purity samples. Finally and perhaps
most important for applications, Bi$_2$Se$_3$ has a large
band gap of around 0.3 eV (3600 K). This indicates that in its
high-purity form Bi$_2$Se$_3$ can exhibit topological insulator
behavior at room temperature and greatly increases
the potential for applications, which we discuss in greater depth later in the review.

\begin{figure*}
\includegraphics[width=17cm]{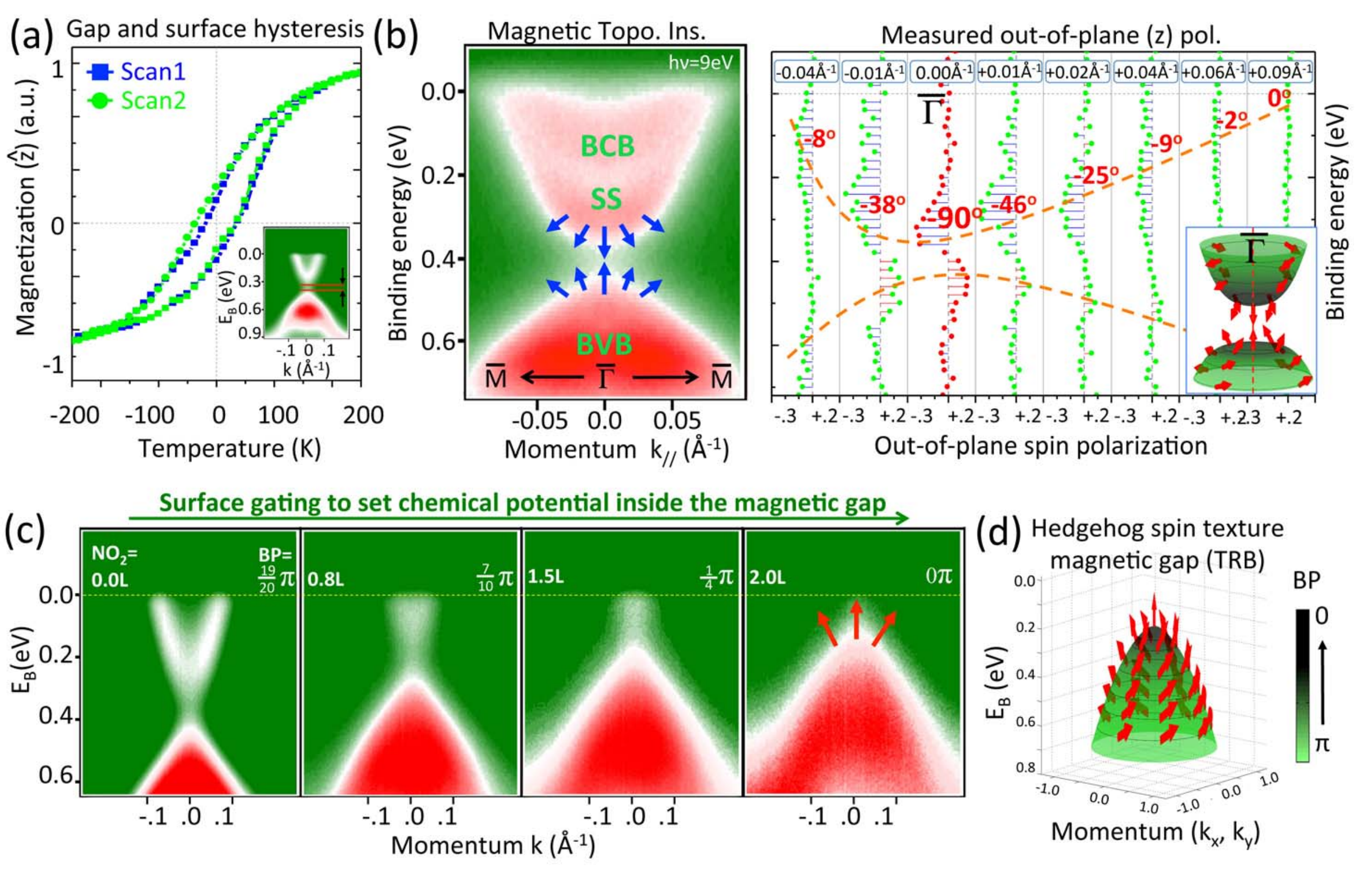}
\caption{\textbf{Hedgehog spin texture and Berry's phase tuning in a magnetic topological insulator.} (a) Magnetization measurements using magnetic circular dichroism shows out-of-plane ferromagnetic character of the Mn-Bi$_2$Se$_3$ MBE film surface through the observed hysteretic response. The inset shows the ARPES observed gap at the Dirac point in the Mn(2.5\%)-Bi$_2$Se$_3$ film sample. (b) Spin-integrated and spin-resolved measurements on a representative piece of Mn(2.5\%)-Bi$_2$Se$_3$ film sample using 9 eV photons. Left: Spin-integrated ARPES dispersion map. The blue arrows represent the spin texture configuration in close vicinity of the gap revealed by our spin-resolved measurements. Right, Measured out-of-plane spin polarization as a function of binding energy at different momentum values. The momentum value of each spin polarization curve is noted on the top. The polar angles ($\theta$) of the spin polarization vectors obtained from these measurements are also noted. The $90^{\circ}$ polar angle observed at $\bar{\Gamma}$ point suggests that the spin vector at $\bar{\Gamma}$ is along the vertical direction. The spin behavior at $\bar{\Gamma}$ and its surrounding momentum space reveals a hedgehog-like spin configuration for each Dirac band separated by the gap. Inset shows a schematic of the revealed hedgehog-like spin texture. (c) Measured surface state dispersion upon \textit{in situ} NO$_2$ surface adsorption on the Mn-Bi$_2$Se$_3$ surface. The NO$_2$ dosage in the unit of Langmuir ($1\textrm{L}=1\times10^{-6}$ torr${\cdot}$sec) and the tunable Berry's phase (BP) associated with the topological surface state are noted on the top-left and top-right corners of the panels, respectively. The red arrows depict the time-reversal breaking out-of-plane spin texture at the gap edge based on the experimental data. (d) The time-reversal breaking spin texture features a singular hedgehog-like configuration when the chemical potential is tuned to lie within the magnetic gap, corresponding to the experimental condition presented in the last panel in panel (c). [Adapted from S.-Y. Xu $et$ $al.$, \textit{Nature Physics} \textbf{8}, 616 (2012). \cite{Hedgehog}].}
\end{figure*}

\section{Identifying the origin of 3D topological order via a bulk band gap inversion transition}

We investigated the quantum origin of topological order in the Bi$_{1-x}$Sb$_x$ and Bi$_2$Se$_3$
classes of materials. It has been theoretically speculated that the
novel topological order in Bi$_{1-x}$Sb$_x$ originates from the parities of the
electrons in pure Sb and not Bi \cite{11,Lenoir}. It was also noted \cite{20} that
the origin of the topological effects can only be tested by
measuring the spin-texture of the Sb surface, which has not been
measured. Based on quantum oscillation and magneto-optical studies,
the bulk band structure of Sb is known to evolve from that of
insulating Bi$_{1-x}$Sb$_x$ through the hole-like band at H rising
above $E_\textrm{F}$ and the electron-like band at $L$ sinking below $E_\textrm{F}$ \cite{Lenoir}. The relative energy ordering of the $L_a$ and $L_s$ states in
Sb again determines whether the surface state pair emerging from
$\bar{\Gamma}$ switches partners [Fig.\ref{Sb_Fig2}(A)] or not [Fig.\ref{Sb_Fig2}(B)]
between $\bar{\Gamma}$ and $\bar{M}$, and in turn determines whether
they support a non-zero Berry's phase.

\begin{figure*}
\includegraphics[scale=.75,clip=true, viewport=-0.0in 0in 6.8in 6.8in]{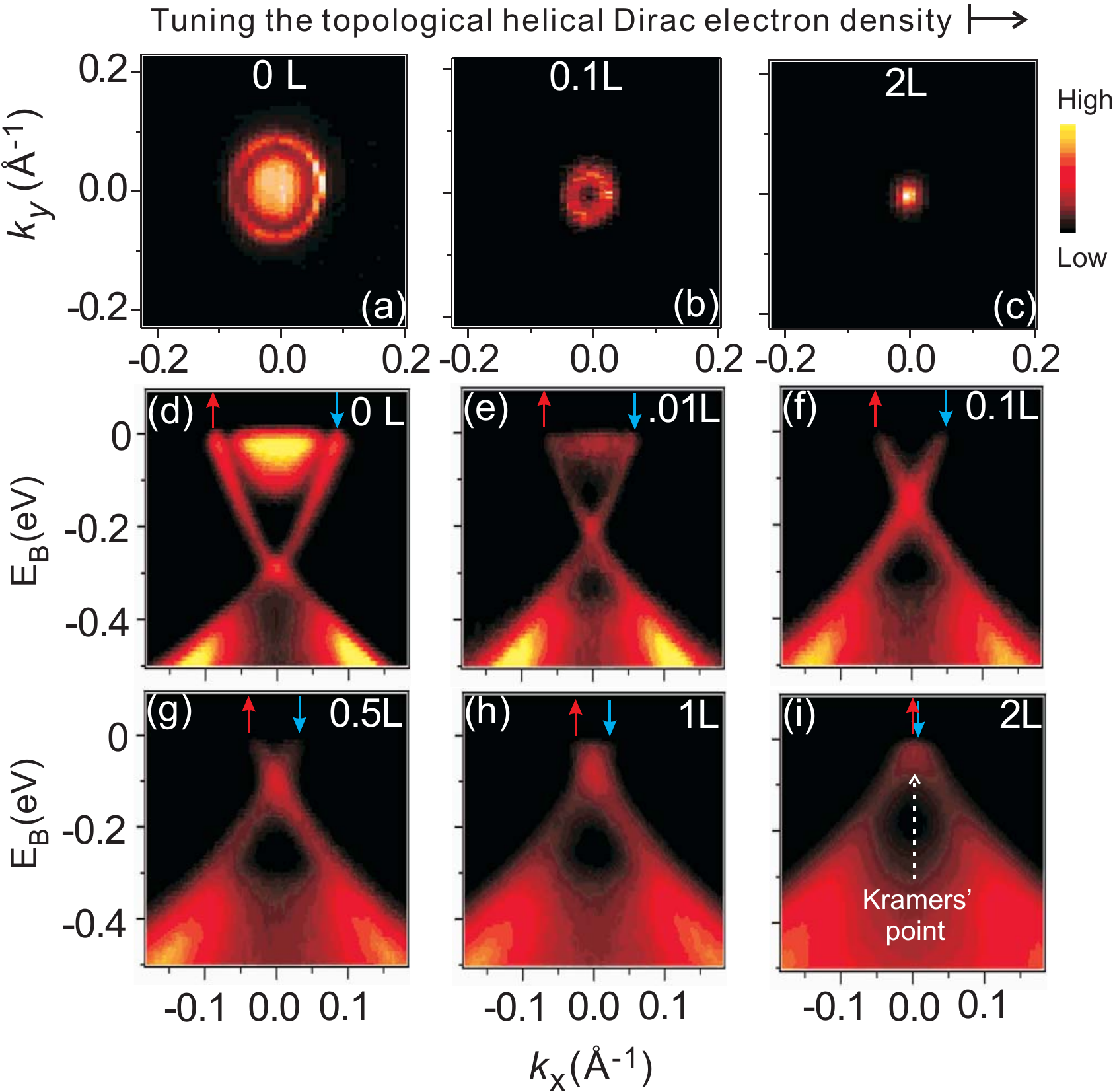}
\caption{\label{fig:Fig3} \textbf{Surface Gating : Tuning the density of helical Dirac electrons to the spin-degenerate Kramers point and topological transport regime.}
(a) A high resolution ARPES mapping of the surface Fermi surface
(FS) near $\bar{\Gamma}$ of Bi$_{2-\delta}$Ca$_{\delta}$Se$_3$
(111). The diffuse intensity within the ring originates from the
bulk-surface resonance state \cite{15}. (b) The FS after 0.1
Langmuir (L) of NO$_2$ is dosed, showing that the resonance state is
removed. (c) The FS after a 2 L dosage, which achieves the Dirac
charge neutrality point. (d) High resolution ARPES surface band
dispersions through after an NO$_2$ dosage of 0 L, (e) 0.01 L, (f)
0.1 L, (g) 0.5 L, (h) 1 L and (i) 2 L. The arrows denote the spin
polarization of the bands. We note that due to an increasing level
of surface disorder with NO$_2$ adsorption, the measured spectra
become progressively more diffuse and the total photoemission
intensity from the buried Bi$_{2-\delta}$Ca$_{\delta}$Se$_3$ surface
is gradually reduced. [Adapted from D. Hsieh $et$ $al.$, \textit{Nature} \textbf{460}, 1101 (2009). \cite{Nature_2009}].}
\end{figure*}

In a conventional spin-orbit metal such as gold, a free-electron
like surface state is split into two parabolic spin-polarized
sub-bands that are shifted in $\vec{k}$-space relative to each other \cite{18}. Two concentric spin-polarized Fermi surfaces are created, one
having an opposite sense of in-plane spin rotation from the other,
that enclose $\bar{\Gamma}$. Such a Fermi surface arrangement, like
the schematic shown in figure \ref{Sb_Fig2}(B), does not support a non-zero
Berry's phase because the $\vec{k}_T$ are enclosed an even number of
times (2 for most known materials).

\begin{figure*}
\includegraphics[scale=0.84,clip=true, viewport=0.0in 0.0in 7.0in 5.8in]{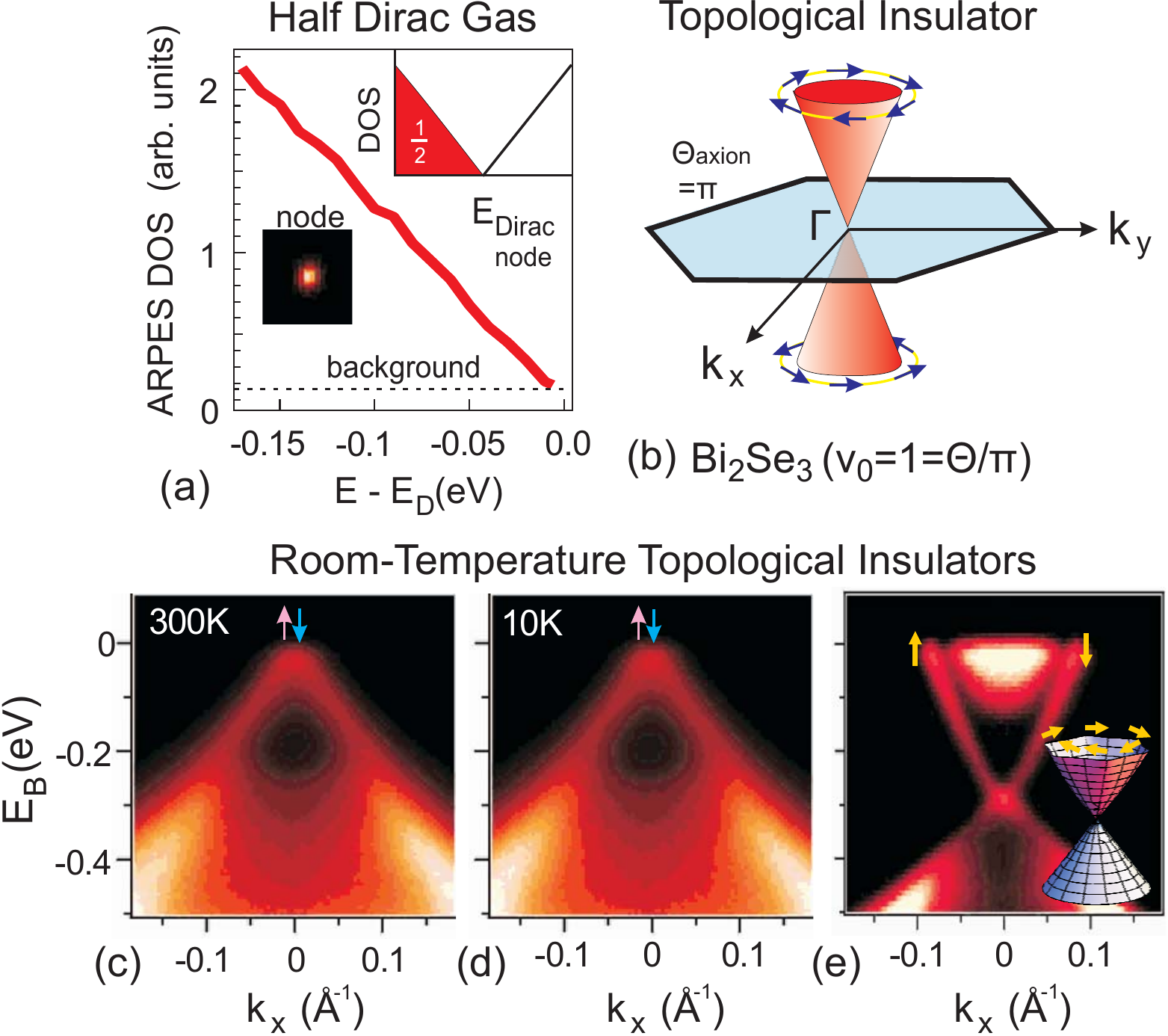}
\caption{\label{RMP_Fig13} \textbf{Observation of room temperature (300K) topological order without applied magnetic field in Bi$_2$Se$_3$:} (a)
Crystal momentum integrated ARPES data near Fermi level
exhibit linear fall-off of density of states, which, combined
with the spin-resolved nature of the states, suggest that a half
Fermi gas is realized on the topological surfaces. (b) Spin texture
map based on spin-ARPES data suggest that the spin-chirality changes sign across the Dirac point. (c) The Dirac
node remains well defined up a temperature of 300 K suggesting
the stability of topological effects up to the room temperature.
(d) The Dirac cone measured at a temperature of 10 K. (e) Full Dirac cone. [Adapted from D. Hsieh $et$ $al.$, \textit{Nature} \textbf{460}, 1101 (2009). \cite{Nature_2009}].}
\end{figure*}

However, for Sb, this is not the case. Figure \ref{Sb_Fig2}(C) shows a
spin-integrated ARPES intensity map of Sb(111) from $\bar{\Gamma}$
to $\bar{M}$. By performing a systematic incident photon energy
dependence study of such spectra, previously unavailable with He
lamp sources \cite{24}, it is possible to identify two V-shaped surface
states (SS) centered at $\bar{\Gamma}$, a bulk state located near
$k_x$ = $-$0.25 \AA$^{-1}$ and resonance states centered about $k_x$
= 0.25 \AA$^{-1}$ and $\bar{M}$ that are hybrid states formed by surface
and bulk states \cite{19}. An examination of the ARPES
intensity map of the Sb(111) surface and resonance states at $E_\textrm{F}$
[Fig.\ref{Sb_Fig2}(E)] reveals that the central surface FS enclosing
$\bar{\Gamma}$ is formed by the inner V-shaped SS only. The outer
V-shaped SS on the other hand forms part of a tear-drop shaped FS
that does \textit{not} enclose $\bar{\Gamma}$, unlike the case in
gold. This tear-drop shaped FS is formed partly by the outer
V-shaped SS and partly by the hole-like resonance state. The
electron-like resonance state FS enclosing $\bar{M}$ does not affect the
determination of $\nu_0$ because it must be doubly spin degenerate
. Such a FS geometry [Fig.\ref{Sb_Fig2}(G)] suggests that the
V-shaped SS pair may undergo a partner switching behavior expected
in Fig.\ref{Sb_Fig2}(A). This behavior is most clearly seen in a cut taken along
the $\bar{\Gamma}-\bar{K}$ direction since the top of the bulk valence
band is well below $E_\textrm{F}$ [Fig.\ref{Sb_Fig2}(F)] showing only the inner V-shaped
SS crossing $E_\textrm{F}$ while the outer V-shaped SS bends back towards the
bulk valence band near $k_x$ = 0.1 \AA$^{-1}$ before reaching $E_\textrm{F}$.
The additional support for this band dispersion behavior comes from
tight binding surface calculations on Sb [Fig.\ref{Sb_Fig2}(D)], which closely
match with experimental data below $E_\textrm{F}$. Our observation of a
single surface band forming a FS enclosing $\bar{\Gamma}$ suggests
that pure Sb is likely described by $\nu_0$ = 1, and that its
surface may support a Berry's phase.

Confirmation of a surface $\pi$ Berry's phase rests critically on a
measurement of the relative spin orientations (up or down) of the SS
bands near $\bar{\Gamma}$ so that the partner switching is indeed
realized, which cannot be done without spin resolution. Spin
resolution was achieved using a Mott polarimeter that measures two
orthogonal spin components of a photoemitted electron \cite{27,28}. These
two components are along the $y'$ and $z'$ directions of the Mott
coordinate frame, which lie predominantly in and out of the sample
(111) plane respectively. Each of these two directions represents a
normal to a scattering plane defined by the photoelectron incidence
direction on a gold foil and two electron detectors mounted on
either side (left and right) [Fig.\ref{Sb_Fig3}(A)]. Strong spin-orbit coupling
of atomic gold is known to create an asymmetry in the scattering of
a photoelectron off the gold foil that depends on its spin component
normal to the scattering plane \cite{28}. This leads to an asymmetry
between the left intensity ($I^L_{y',z'}$) and right intensity
($I^R_{y',z'}$) given by $A_{y',z'} =
(I^L_{y',z'}-I^R_{y',z'})/(I^L_{y',z'}+I^R_{y',z'})$, which is
related to the spin polarization $P_{y',z'} = (1/S_{eff})\times
A_{y',z'}$ through the Sherman function $S_{eff}$ = 0.085 \cite{27,28}.
Spin-resolved momentum distribution curve data sets of the SS bands
along the $-\bar{M}-\bar{\Gamma}-\bar{M}$ cut at $E_\textrm{B}$ = $-$30 meV
[Fig.\ref{Sb_Fig3}(B)] are shown for maximal intensity. Figure \ref{Sb_Fig3}(D) displays
both $y'$ and $z'$ polarization components along this cut, showing
clear evidence that the bands are spin polarized, with spins
pointing largely in the (111) plane. In order to estimate the full
3D spin polarization vectors from a two component measurement (which
is not required to prove the partner switching or the Berry's
phase), we fit a model polarization curve to our data following the
recent demonstration in Ref-\cite{26}, which takes the polarization
directions associated with each momentum distribution curve peak
[Fig.\ref{Sb_Fig3}(C)] as input parameters, with the constraint that each
polarization vector has length one (in angular momentum units of
$\hbar$/2). Our fitted polarization vectors are displayed in the
sample ($x,y,z$) coordinate frame [Fig.\ref{Sb_Fig3}(F)], from which we derive
the spin-resolved momentum distribution curves for the spin
components parallel ($I_y^{\uparrow}$) and anti-parallel
($I_y^{\downarrow}$) to the $y$ direction as shown
in figure \ref{Sb_Fig3}(E). There is a clear difference in $I_y^{\uparrow}$ and
$I_y^{\downarrow}$ at each of the four momentum distribution curve
peaks indicating that the surface state bands are spin polarized
[Fig.\ref{Sb_Fig3}(E)], which is possible to conclude even without a full 3D
fitting. Each of the pairs $l2/l1$ and $r1/r2$ have opposite spin,
consistent with the behavior of a spin split pair, and the spin
polarization of these bands are reversed on either side of
$\bar{\Gamma}$ in accordance with the system being time reversal
symmetric $[E(\vec{k},\uparrow) = E(-\vec{k},\downarrow)]$
[Fig.\ref{Sb_Fig3}(F)]. The measured spin texture of the Sb(111) surface states
(Fig.\ref{Sb_Fig3}), together with the connectivity of the surface bands
(Fig.\ref{Sb_Fig2}), uniquely determines its belonging to the $\nu_0$ = 1 class.
Therefore the surface of Sb carries a non-zero ($\pi$) Berry's phase
via the inner V-shaped band and pure Sb can be regarded as the
parent metal of the Bi$_{1-x}$Sb$_x$ topological insulator class, in
other words, the topological order originates from the Sb wave
functions. A recent work \cite{Xu} has demonstrated a topological quantum phase transition as a function of chemical composition from a non-inverted to an inverted semiconductor as a clear example of the origin of topological order (Fig.\ref{QPT}).

Our spin polarized measurement methods (Fig.\ref{Sb_Fig1} and \ref{Sb_Fig3}) uncover a new
type of topological quantum number $n_M$, which is not related to the time-reversal symmetry, but a consequence of the mirror symmetries of a crystalline system \cite{20}. Topological band theory suggests
that the bulk electronic states in the mirror ($k_y$ = 0) plane can
be classified in terms of a number $n_M$ (=integer) \cite{20}, which defines both the number and the chirality of the spin helical edge states propagating along the mirror planes of a crystal.

For example, we now determine the value of $n_M$ of antimony surface states from our data. From figure \ref{Sb_Fig1}, it is seen that a single (one) surface band, which switches partners at $\bar{M}$, connects the bulk valence and conduction bands, so $|n_M|$ = 1 . The sign of $n_M$ is related to the direction of the spin polarization $\langle \vec{P} \rangle$ of this band \cite{20}, which is constrained by mirror symmetry to point along $\pm\hat{y}$. Since the central electron-like FS enclosing $\bar{\Gamma}$ intersects six mirror invariant points [see Fig.\ref{Sb_Fig3}(B)], the sign of $n_M$ distinguishes two distinct types of handedness for this spin polarized FS. Figures \ref{Sb_Fig1}(F) and \ref{Sb_Fig3} show that for both Bi$_{1-x}$Sb$_x$ and Sb, the surface band that forms this electron pocket has $\langle \vec{P} \rangle \propto -\hat{y}$ along the $k_x$ direction, suggesting a left-handed rotation sense for the spins around this central FS thus $n_M$ = $-1$. We note that similar analysis regarding the mirror symmetry and mirror eigenvalues $n_M=-1$ can be applied to the single Dirac cone surface states in the Bi$_2$Se$_3$ material class.

As a matter of fact, a nonzero (nontrivial) topological mirror number does not require a nonzero Z$_2$ topological number. Or in other word, there is no necessary correlation between a mirror symmetry protected topological order and a time-reversal symmetry protected topological order. However, since most of the Z$_2$ topological insulators (Bi$_{1-x}$Sb$_x$, Bi$_2$Se$_3$, Bi$_2$Te$_3$ and etc.) also possess mirror symmetries in their crystalline form, thus topological mirror order $n_M$ typically coexists with the strong Z$_2$ topological order, and only manifests itself as the chirality or the handness of the spin texture. One possible way to isolate the mirror topological order from the Z$_2$ order is to work with systems that feature an even number of bulk band inversions. This approach naturally exclude a nontrivial Z$_2$ order which strictly requires an odd number of band inversions.  More importantly, if the locations of the band inversions coincide with the mirror planes in momentum space, it will lead to a topologically nontrivial phase protected by the mirror symmetries of the crystalline system that is irrelevant to the time-reversal symmetry protection and the Z$_2$ (Kane-Mele) topological order. Such exotic new phase of topological order, noted as topological crystalline insulator \cite{Liang PRL TCI} protected by space group mirror symmetries, has very recently been theoretically predicted and experimentally identified in the Pb$_{1-x}$Sn$_x$Te(Se) alloy systems \cite{Liang NC SnTe, TCI Hasan, TCI Story, TCIAndo}. An anomalous $n_M=-2$ topological mirror number in Pb$_{1-x}$Sn$_x$Te, distinct from the $n_M=-1$ case observed in the Z$_2$ topological insulators, has also been experimentally determined using spin-resolved ARPES measurements as shown in Ref. \cite{TCI Hasan}. Moreover, the mirror symmetry can be generalized to other space group symmetries, leading to a large number of distinct topological crystalline insulators awaited to be discovered, some of which are predicted to exhibit nontrivial crystalline order even without spin-orbit coupling as well as topological crystalline surface states in non-Dirac (e.g. quadratic) fermion forms \cite{Liang PRL TCI}.

In 3D Z$_2$ (Kane-Mele) TIs, the spin-resolved experimental measurements shown above reveal an intimate and straightforward connection between the topological numbers ($\nu_0$, $n_M$) and the nontrivial Berry's phase. The $\nu_0$ determines whether the surface electrons support a non-trivial Berry's phase. The 2D Berry's phase is a critical signature of topological order and is not realizable in isolated 2D electron systems, nor on the surfaces of conventional spin-orbit or exchange coupled magnetic materials. A non-zero Berry's phase is known, theoretically, to protect an electron system against the almost universal weak-localization behavior in their low temperature transport \cite{11,13} and is expected to form the key element for fault-tolerant computation schemes \cite{13,29}, because the Berry's phase is a geometrical agent or mechanism for protection against quantum decoherence \cite{30}. Its remarkable realization on the Bi$_{1-x}$Sb$_x$ surface represents an unprecedented example of a 2D $\pi$ Berry's phase, and opens the possibility for building realistic prototype systems to test quantum computing modules. In general, our results demonstrate that spin-ARPES is a powerful probe of 3D topological order, which opens up a new search front for topological materials for novel spin-devices and fault-tolerant quantum computing.

\section{Topological protection and tunability of the surface states of a 3D topological insulator}

The predicted topological protection of the surface states of Sb
implies that their metallicity cannot be destroyed by weak time
reversal symmetric perturbations. In order to test the robustness of
the measured gapless surface states of Sb, we introduce such a
perturbation by randomly substituting Bi into the Sb crystal matrix. Another motivation for performing such an experiment
is that the formalism developed by Fu and Kane \cite{11} to calculate the
Z$_2$ topological invariants relies on inversion symmetry being
present in the bulk crystal, which they assumed to hold true even in
the random alloy Bi$_{1-x}$Sb$_x$. However, this formalism is simply
a device for simplifying the calculation and the non-trivial
$\nu_0=1$ topological class of Bi$_{1-x}$Sb$_x$ is predicted to hold
true even in the absence of inversion symmetry in the bulk crystal
\cite{11}. Therefore introducing light Bi substitutional disorder into
the Sb matrix is also a method to examine the effects of alloying
disorder and possible breakdown of bulk inversion symmetry on the
surface states of Sb(111). We have performed spin-integrated ARPES
measurements on single crystals of the random alloy
Sb$_{0.9}$Bi$_{0.1}$. Figure~\ref{fig:Sb_FigS5} shows that both the
surface band dispersion along $\bar{\Gamma}$-$\bar{M}$ as well as the
surface state Fermi surface retain the same form as that observed in
Sb(111), and therefore the `topological metal' surface state of
Sb(111) fully survives the alloy disorder. Since Bi alloying is seen
to only affect the band structure of Sb weakly, it is reasonable to
assume that the topological order is preserved between Sb and
Bi$_{0.91}$Sb$_{0.09}$
as we observed.

In a simpler fashion compared to Bi$_{1-x}$Sb$_x$, the topological insulator behavior in Bi$_2$Se$_3$
is associated with a single band inversion at the surface Brillouin zone center. Owing to its larger bandgap compared with Bi$_{1-x}$Sb$_x$, ARPES has shown that its topological properties are preserved at room temperature \cite{Nature_2009}. Two defining properties of topological insulators,
spin-momentum locking of surface states and $\pi$ Berry
phase, can be clearly demonstrated in the Bi$_2$Se$_3$ series.
The surface states are protected by time-reversal symmetry symmetry, which implies that the surface Dirac node should be robust in the presence of nonmagnetic disorder. On the other hand, a gap at the Dirac point is theoretically expected to open in the presence of magnetism along the out-of-plane direction perpendicular to the sample surface. Here magnetism is required to break time-reversal symmetry whereas its out-of-plane direction is critical to break additional protections by crystalline-symmetries  \cite{Liang PRL TCI}. Unlike in theory, magnetism in real topological insulator materials exhibits complex phenomenology \cite{Wray2, Hor PRB BiMnTe, Cava Fe, Salman Fe, vdW, Gap, Fe XMCD, Chen Science Fe}. Experimental studies of a magnetic gap especially on the quantitative level is challenging because many other physical or chemical changes are found to also lead to extrinsic gap-like feature at the Dirac point \cite{Ando QPT, Haim Nature physics BiSe, vdW, Gap}. We utilize spin-resolved angle-resolved photoemission spectroscopy to measure the momentum space spin configurations in systematically magnetically (Mn) doped, non-magnetically (Zn) doped, and ultra-thin quantum coherent topological insulator films \cite{Hedgehog}. Fig. 14b shows the out-of-plane spin polarization ($P_z$) measurements of the electronic states in the vicinity of the Dirac point gap of a Mn(2.5\%)-Bi$_2$Se$_3$ sample. The surface electrons at the time-reversal invariant $\bar{\Gamma}$ point (red curve in Fig. 14b) are clearly observed to be spin polarized in the out-of-plane direction. The opposite sign of $P_z$ for the upper and lower Dirac band shows that the Dirac point spin degeneracy is indeed lifted up ($\textrm{E}(k_{//}=0,\uparrow){\neq}\textrm{E}(k_{//}=0,\downarrow)$), which manifestly breaks the time-reversal symmetry on the surface of our Mn(2.5\%)-Bi$_2$Se$_3$ samples. Systematic spin-resolved measurements as a function of binding energy and momentum reveal a Hedgehog-like spin texture (inset of Fig. 14b). Such exotic spin groundstate in a magnetic topological insulator enables a tunable Berry's phase on the magnetized topological surface \cite{Hedgehog}, as experimentally demonstrated by our chemical gating via NO$_2$ surface adsorption method shown in Figs. 14c and d.

Many of the interesting theoretical proposals that utilize
topological insulator surfaces require the chemical
potential to lie at or near the surface Dirac point. This is
similar to the case in graphene, where the chemistry of
carbon atoms naturally locates the Fermi level at the
Dirac point. This makes its density of carriers highly
tunable by an applied electrical field and enables applications
of graphene to both basic science and microelectronics.
The surface Fermi level of a topological insulator
depends on the detailed electrostatics of the surface
and is not necessarily at the Dirac point. Moreover, for
naturally grown Bi$_2$Se$_3$ the bulk Fermi energy is not
even in the gap. The observed n-type behavior is believed
to be caused Se vacancies. By appropriate chemical
modifications, however, the Fermi energy of both the
bulk and the surface can be controlled. This allowed \cite{Nature_2009} to reach the sweet
spot in which the surface Fermi energy is tuned to the
Dirac point (Fig.\ref{fig:Fig3}). This was achieved by doping bulk
with a small concentration of Ca, which compensates the
Se vacancies, to place the Fermi level within the bulk
band gap. The surface was the hole doped by exposing the
surface to NO$_2$ gas to place the Fermi level at the Dirac
point, and has been shown to be effective even at room temperature (Fig.\ref{RMP_Fig13}). These results collectively show how ARPES can be used to study the topological protection and tunability properties of the 2D surface of a 3D topological insulator.

These techniques to identify, characterize and manipulate the topological bulk and surface states of 3D topological insulators have opened the way for performing surface sensitive transport measurements on bulk crystals \cite{Qu,Analytis}, exfoliated nano-devices \cite{Hadar1,Chen1} and thin films \cite{FuChun,Yayu,Samarth} as well as surface sensitive optical measurements \cite{Hsieh_SHG, Kerr}.

\section{Future directions: Topological Superconductors and Topological Crystalline Insulators}

\begin{figure*}
\includegraphics[scale=0.3,clip=true, viewport=0.0in 0.0in 20.5in 13.5in]{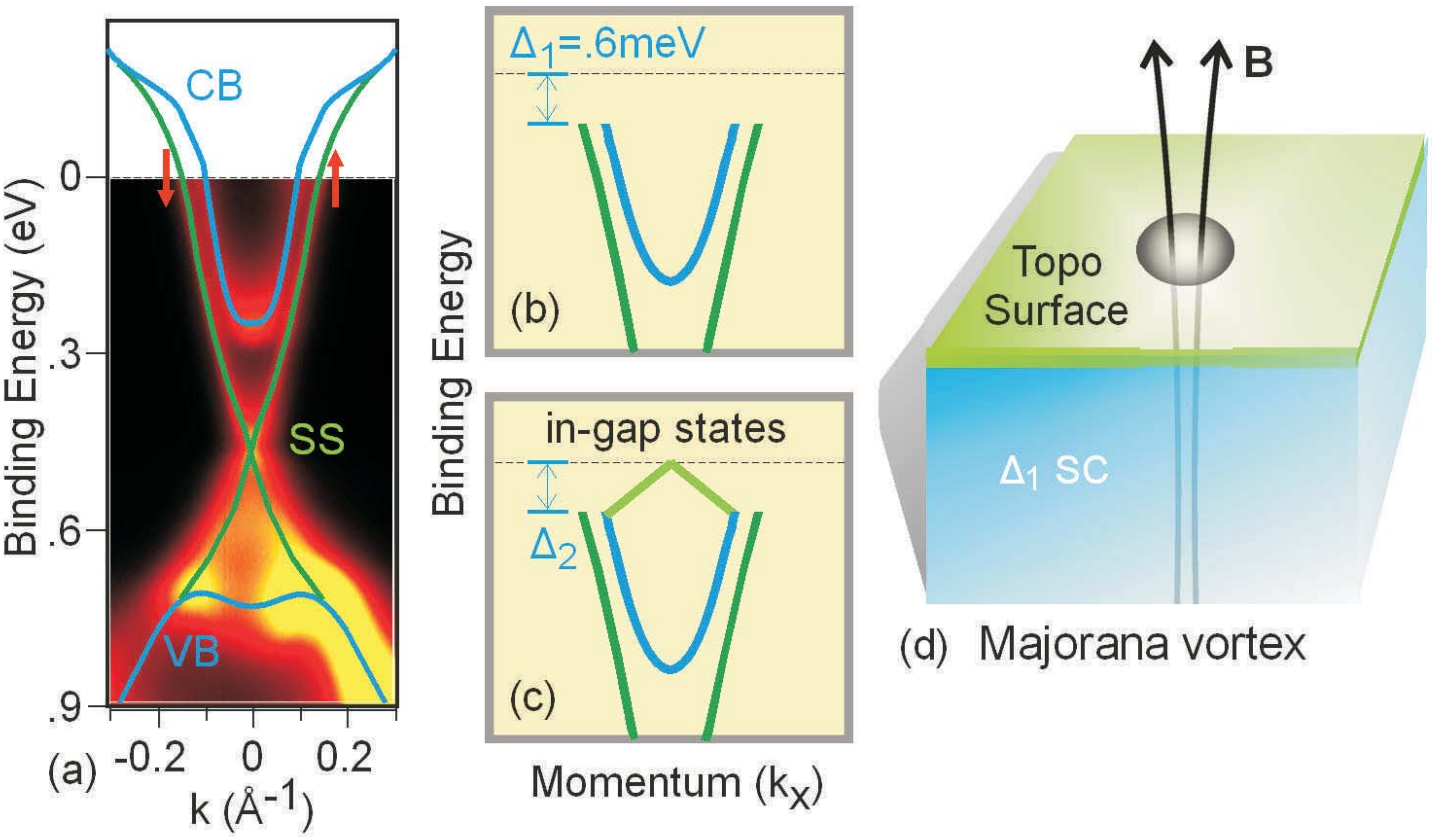}
\caption{\label{Figure13} \textbf{A Majorana platform.} (a) Topologically protected surface states cross the Fermi level before merging with the
bulk valence and conduction bands in a lightly doped topological insulator. (b) If the superconducting wavefunction has even
parity, the surface states will be gapped by the proximity effect, and vortices on the crystal surface will host braidable Majorana
fermions. (c) If superconducting parity is odd, the material will be a so-called topological superconductor, and new states
will appear below T$_c$ to span the bulk superconducting gap. (d) Majorana fermion surface vortices are found at the end of
bulk vortex lines and could be manipulated for quantum computation if superconducting pairing is even. [Adapted from L. Wray $et$ $al$., \textit{Nature Phys.} \textbf{6}, 855 (2010). \cite{Wray1}]}
\end{figure*}

Recent measurements \cite{Wray1} show that surface
instabilities cause the spin-helical topological insulator
band structure of Bi$_2$Se$_3$ to remain well defined and non-
degenerate with bulk electronic states at the Fermi level
of optimally doped superconducting Cu$_{0.12}$Bi$_2$Se$_3$, and
that this is also likely to be the case for superconduct-
ing variants of p-type Bi$_2$Te$_3$. These surface states pro-
vide a highly unusual physical setting in which super-
conductivity cannot take a conventional form, and is expected to realize one of two novel states that have not
been identified elsewhere in nature. If superconducting
pairing has even parity, as is nearly universal among the
known superconducting materials, the surface electrons
will achieve a 2D non-Abelian superconductor state with
non-commutative Majorana fermion vortices that can potentially be manipulated to store quantum information.
Surface vortices will be found at the end of bulk vortex
lines as drawn in Fig.\ref{Figure13}. If superconducting pairing
is odd, the resulting state is a novel state of matter known
as a ``topological superconductor" with Bogoliubov surface quasi-particles present below the superconducting critical temperature of 3.8 K. As drawn in Fig.\ref{Figure13}(c), these low temperature surface states would be gapless,
likely making it impossible to adiabatically manipulate
surface vortices for quantum computation. The unique
physics and applications of the topological superconductor state are distinct from any known material system,
and will be an exciting vista for theoretical and experimental exploration if they are achieved for the first time
in Cu$_x$Bi$_2$Se$_3$.

\begin{figure*}
\includegraphics[width=17cm]{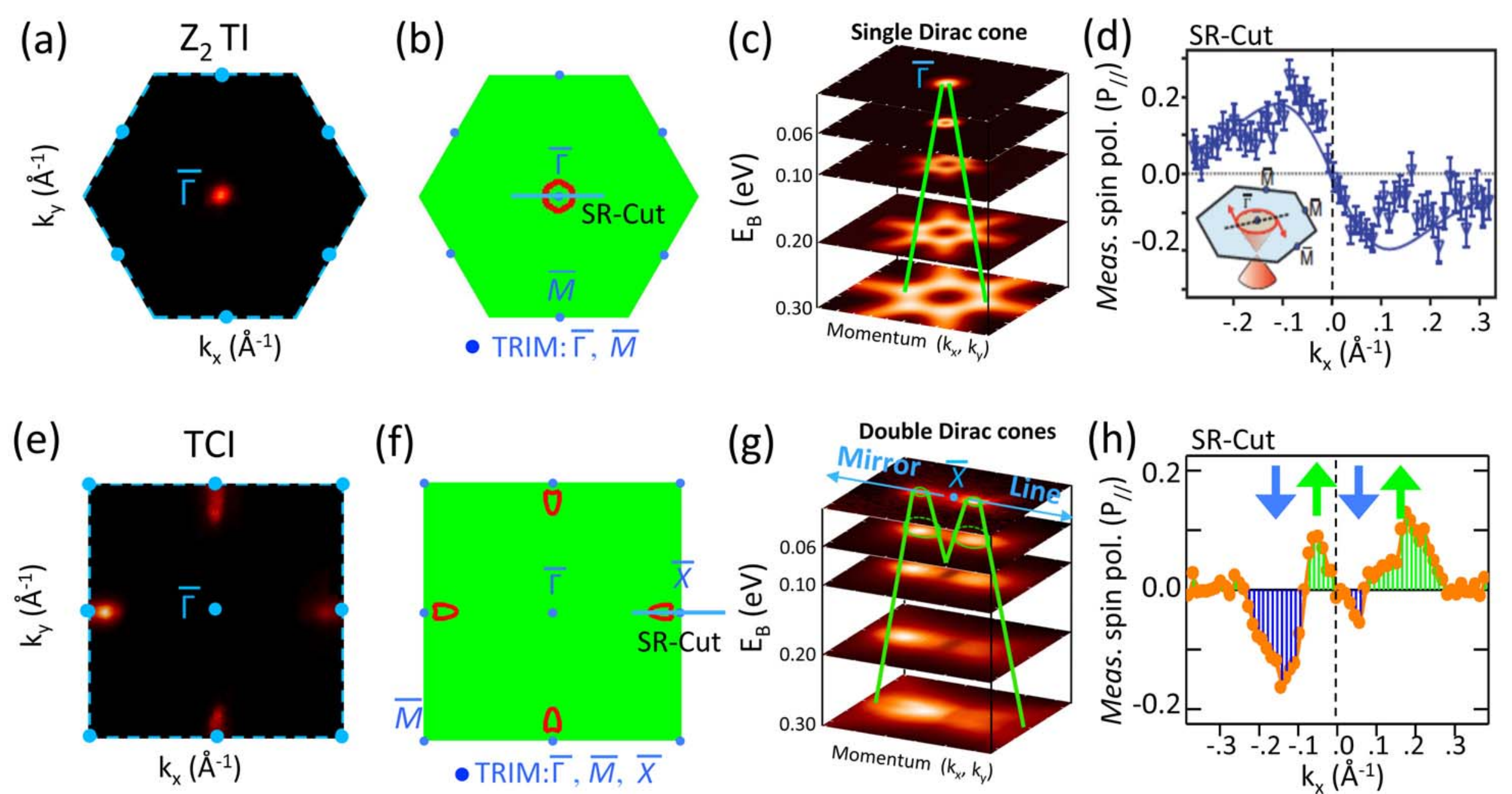}
\caption{\textbf{The topological distinction between Z$_2$ (Kane-Mele) topological insulator and topological crystalline insulator phases.} (a-d) ARPES, spin-resolved ARPES and calculation results of the surface states of a Z$_2$ topological insulator GeBi$_2$Te$_4$ \cite{Ternary arXiv}, an analog to Bi$_2$Se$_3$ \cite{Xia}. (a) ARPES measured Fermi surface with the chemical potential tuned near the surface Dirac point. (b) First-principles calculated iso-energetic contour of the surface states near the Dirac point. The solid blue line shows the momentum-space cut used for spin-resolved measurements. Right: A stack of ARPES iso-energetic contours near the $\bar{\Gamma}$ point of the surface BZ. (d) Measured spin polarization of Bi$_2$Se$_3$, in which a helical spin texture is revealed. (e-h)  ARPES and spin-resolved ARPES measurements on the Pb$_{0.6}$Sn$_{0.4}$Te ($x=0.4$) samples and band calculation results on the end compound SnTe \cite{Liang NC SnTe}. (e) ARPES measured Fermi surface map of Pb$_{0.6}$Sn$_{0.4}$Te. (f) First-principles calculated iso-energetic contour of SnTe surface states near the Dirac point. The solid blue line shows the momentum-space cut near the surface BZ edge center $\bar{X}$ point, which is used for spin-resolved measurements shown in panel (h). (g) A stack of ARPES iso-energetic contours near the $\bar{X}$ point of the surface BZ, revealing the double Dirac cone contours near each $\bar{X}$ point on the surface of Pb$_{0.6}$Sn$_{0.4}$Te. (h) Measured spin polarization of Pb$_{0.6}$Sn$_{0.4}$Te near the native Fermi energy along the momentum space cut defined in panel (f), in which two spin helical  Dirac cones are observed near an $\bar{X}$ point. [Adapted from S.-Y. Xu $et$ $al.$, \textit{Nature Comm.} \textbf{3}, 1192 (2012). \cite{TCI Hasan}. Without spin measurements it is not possible to identify TCI surface states hence no proof of the TCI phase was possible prior to the measurements in Ref-\cite{TCI Hasan}.]}
\end{figure*}

Another interesting frontier involves a newly discovered topological phase of matter, namely a topological crystalline insulator where space group symmetries replace the role of time-reversal symmetry in a Z$_2$ topological insulator \cite{Liang PRL TCI, Liang NC SnTe, TCI Hasan, ando, TCIAndo, TCI Story}. In Ref.~\cite{TCI Hasan}, we experimentally investigate the possibility of a mirror symmetry protected topological phase transition in the Pb$_{1-x}$Sn$_x$Te alloy system, which has long been known to contain an even number of band inversions based on band theory. We show that at a composition below the theoretically predicted band inversion, the system is fully gapped \cite{TCI Hasan}, whereas in the inverted regime, the surface exhibits even number of spin-polarized Dirac cone states revealing mirror-protected topological order distinct from that observed in Bi$_{1-x}$Sb$_x$ and Bi$_2$Se$_3$. Fig. 18 presents a comparison between the Pb$_{0.6}$Sn$_{0.4}$Te and a single Dirac cone Z$_2$ topological insulator (TI) GeBi$_2$Te$_4$ \cite{Ternary arXiv}. In clear contrast to GeBi$_2$Te$_4$, which features a single spin helical Dirac cone enclosing the time-reversal invariant (Kramers') $\bar{\Gamma}$ point [Fig. 18(a-d)], none of the surface states of Pb$_{0.6}$Sn$_{0.4}$Te is observed to enclose any of the time-reversal invariant momentum, demonstrating their irrelevance to the time-reversal symmetry related protection. On the other hand, all of the Pb$_{0.6}$Sn$_{0.4}$Te surface states are found to locate along the two independent momentum space mirror line ($\bar{\Gamma}-\bar{X}-\bar{\Gamma}$) directions, revealing  their topological protection as a consequence of the crystalline mirror symmetries of the Pb$_{1-x}$Sn$_{x}$Te system \cite{TCI Hasan}. The irrelevance to time-reversal symmetry observed in Ref.~\cite{TCI Hasan} suggests the potential to realize magnetic yet topologically protected surface states on the Pb$_{1-x}$Sn$_{x}$Te surface, which is fundamentally not possible in the Z$_2$ topological insulator systems. In general, our discovery of a mirror symmetry protected topological phase in the Pb$_{1-x}$Sn$_{x}$Te system opens the door for a wide range of exotic new physics, such as Lifshitz transition and Fermi surface Fractionalization based on topological surface states \cite{Liang NC SnTe}, topological antiferromagnetic insulator \cite{PbMnTe, AFM} and topological crystalline superconductor \cite{In-SnTe Ando, TCS} via bulk doping or proximity interfacing, as well as new topological crystalline orders protected by other point group symmetries even without spin-orbit interactions \cite{Liang PRL TCI}.

\bigskip
\textbf{Acknowledgement}
The authors acknowledge A. Bansil, H. Lin, C. L. Kane, A. V. Fedorov, J.H. Dil, J. Osterwalder and D. Qian for collaboration; R.J. Cava, N. Samarth and F.C. Chou groups for samples; and U.S. DOE/BES DE-FG-02-05ER46200 for primary support. Authors also acknowledge invaluable technical support from many scientists at national and international facilities including Advanced Light Source (LBNL), Stanford Synchrotron facility (SSRL/SLAC), Swiss Light Source (SLS), Wisconsin Synchrotron facility (SRC) and MaxLab (Lund/Sweden). M.Z.H. acknowledges partial support for sample growth through ACI Fellowship and visiting scientist support from Lawrence Berkeley National Laboratory and additional support from A.P. Sloan Foundation and Princeton University.

\end{document}